\documentclass[twocolumn,aps,pre,showpacs,floatfix]{revtex4}
\usepackage{graphics}
\usepackage{graphicx}
\usepackage{amsmath}
\usepackage{amssymb}
\usepackage{color}
\usepackage{ulem}
\usepackage[latin1]{inputenc}

\begin{document}
\title{Mean-Field Approximation for Spacing Distribution Functions in Classical Systems}
\author{Diego Luis Gonz\'alez$^{1,*}$,
Alberto Pimpinelli$^{1,2,\dagger}$,
and T.L. Einstein$^{1,*}$
}
\email[]{dgonzal2@umd.edu}
\email[$^\dagger$]{apimpin1@umd.edu}
\email[$^{**}$]{einstein@umd.edu}
\affiliation{$^{1}$Department of Physics, University of Maryland, College Park, Maryland 20742-4111  USA \\
$^{2}$French Embassy, Consulate General of France, Houston, Texas 77056 USA}

\date{\today}

\begin{abstract}
We propose a mean-field method to calculate approximately the spacing distribution functions $p^{(n)}(s)$ in 1D classical many-particle systems. We compare our method with two other commonly used methods, the independent interval approximation (IIA) and the extended Wigner surmise (EWS). In our mean-field approach, $p^{(n)}(s)$ is calculated from a set Langevin equations which are decoupled by using a mean-field approximation. We found that in spite of its simplicity, the mean-field approximation provides good results in several systems. We offer many examples in which the three methods mentioned previously give a reasonable description of the statistical behavior of the system. The physical interpretation of each method is also discussed.
\end{abstract}



\maketitle

\section{Introduction}
The spacing distribution functions $p^{\left(n\right)}(s)$ are often used to describe the statistical behavior of many-particle systems in one dimension (1D) \cite{Bogomolny,salsburg,forrester,ben2,doering,ben0,alemany}. By definition, $\hat{p}^{(n)}(S)$ is the probability density that an interval of length $S$ which starts at a particle contains exactly $n$ particles and that the next, the $(n\! +\! 1)^{\rm th}$ particle, is in $[S,S+dS]$. The relative spacing is defined as $s=S/\left\langle S\right\rangle$, where $\left\langle S\right\rangle$ is the average of $S$. However, $p^{(n)}(s)$ are also used in other contexts. For example, in 1D systems with domains, $S$ represents the spacing  between domain boundaries \cite{mettetal,cornell,spirin,derrida2,gonzalez,aarao,Derrida-Godre,gonzalez3}, while in random matrix theory and in quantum systems this variable represents the spacing between adjacent energy eigenvalues \cite{mehta,abul,guy,terras,mejia,kaufman}. In the physics of surfaces, $S$ can be the distance between islands in epitaxial growth models or the terrace width between adjacent steps on vicinal (misoriented) surfaces \cite{pimpinelli,einstein,blackman,pimpinelli1}.

The spacing distribution functions are useful even in non-equilibrium systems having dynamical scaling. One such system is the coalescing random walk (CRW). In CRW all particles execute independent random walks, suffering a fusion reaction $\left(A+A\rightarrow A\right)$ when two particles meet. Clearly in this system the number of particles and $\left\langle S(t)\right\rangle$ are time dependent. Despite this, it is possible to define a scaled spacing distribution according to
\begin{equation}\label{scaling}
p^{\left(n\right)}(s)=\left\langle S\right\rangle
\hat{p}^{\left(n\right)}(s\left\langle S\right\rangle,t)
\,.
\end{equation}
Since $p^{\left(n\right)}(s)$ in Eq.~(\ref{scaling}) does not depend on $t$, it can be compared with the spacing distribution functions of an equilibrium system. More information about the CRW is given in Ref.~\cite{ben}.

In this paper we propose a mean-field model to calculate in an approximate way the spacing distribution functions of classical many-particle systems. Our method is compared with other existing methods commonly used to calculate $p^{(n)}(s)$. These approximate methods are important because they are often used to obtain the spacing distribution functions for systems where the exact analytical solution  either cannot be obtained or, if it is obtained, cannot be handled easily because of its complexity. In Section II, we provide a brief introduction to the spacing distribution functions, giving some important definitions. In Sections III and IV we briefly review the independent interval approximation (IIA) and the extended Wigner surmise (EWS) \cite{EWSGWS}, respectively. (NB, the EWS differs  \cite{EWSGWS} from the so-called generalized Wigner surmise (GWS) \cite{pimpinelli,pimpinelli1,einstein}, which is not used in this paper.)  In Section V we develop two mean-field models. In each case we provide several examples of the use of these methods, analyzing their advantages and limitations. Finally, in Section VI we offer conclusions.

All our numerical data for Dyson's Brownian motion model were generated with a Monte Carlo simulation that uses the standard Metropolis algorithm. We used a lattice with $L=10000$ sites with $N=50$ particles. The statistics take into account over $20000$ realizations.

\section{Basic Definitions}

Consider $N$ particles which can move around a circle of circumference $L$; periodic boundary conditions are imposed, that is, $x_{N+j}=x_j$, where $x_j$ is the position of the $j$-th particle. If the system is in equilibrium at an inverse temperature $\beta$, then its statistical behavior is totally defined by the joint probability distribution $P_{N}\left(x_1,\cdots,x_N;\beta\right)$
\begin{equation}\label{joint}
P_{N}\left(x_1,\cdots,x_N;\beta\right)=\frac{1}{Z_N(L;\beta)}e^{-\beta\,V(x_1,\cdots,x_N)},
\end{equation}
where $V(x_1,\cdots,x_N)$ is the total interaction energy among the $N$ particles, and $Z_N$ is the configurational partition function of the system. Then $P_{N}\left(x_1,\cdots,x_N;\beta\right)$ represents the probability density to find particle 1 in [$x_1$, $x_1+dx_1$], particle 2 in [$x_2$, $x_2+dx_2$], etc.
In practice this joint probability distribution cannot be obtained easily from experiments or numerical simulations because it depends on many variables. Instead, one usually evaluates the spacing distribution functions $\hat{p}^{(n)}(S)$, where $n\ge0$. Each $\hat{p}^{(n)}(S)$ contains reduced information about the system.  In order to have a more complete description of the statistical behavior of the system, it is necessary to know $\hat{p}^{(n)}(S)$ for all $n$.  From them it is possible to calculate the pair correlation function $g(s)$ by using the expression
\begin{equation}\label{paircorr}
g(s)=\sum^{\infty}_{n=0}p^{(n)}(s).
\end{equation}
The sum in Eq.~(\ref{paircorr}) seems to be a formidable task. However, in many cases $g(s)$ quickly relaxes to 1 and therefore, just the first $p^{(n)}(s)$ have to be calculated explicitly. Many thermodynamic quantities can be expressed in terms of the pair correlation function and the interaction potential between particles \cite{chandler,hansen}.

When the $N$ particles interact via a pair potential $v(r)$, the total energy of interaction reduces to
\begin{equation}\label{potential}
V(x_1,\cdots,x_N)=\sum^{N}_{m=1}\sum^{q}_{j=1}v\left(x_{m+j}-x_m\right),
\end{equation}
with $q$ the number of interacting neighbors. For $q=1$ we have nearest-neighbor interactions and for $q=N-1$ each particle interacts with all other particles.  We henceforth denote the latter as ``full-range" interactions, corresponding to infinite range in the thermodynamic limit.  As mentioned previously, the probability density to find the $N$ particles around the positions $x_1,\cdots x_N$ is given by Eq.~(\ref{joint}). As usual, $Z_N$ can be calculated from
\begin{equation}
Z_N(L;\beta)=\int dx_1\cdots dx_N\,\delta\left(\Lambda\right)\,P_{N}\left(x_1,\cdots,x_N;\beta\right),
\end{equation}
where $\Lambda=L-\sum^{N}_{i=1}\left(x_{i+1}-x_i\right)$. Making the change of variables $S_i=x_{i+1}-x_i$, the partition function takes the form
\begin{equation}
Z_N(L;\beta)=\int dS_1\cdots dS_N\,\delta\left(\Lambda\right)\,P_{N}\left(S_1,\cdots\!,S_N;\beta\right),
\end{equation}
now with $\Lambda=L-\sum^{N}_{i=1}S_{i}$. In the same way, the joint probability distribution can be written as
\begin{equation}
P_{N}\left(S_1,\cdots\!,S_N;\beta\right)=\frac{1}{Z_N(L;\beta)}e^{-\beta\,\Omega},
\end{equation}
with
\begin{eqnarray}
\Omega &=& \sum^{N}_{m=1}\left[v(S_m)+v(S_m+S_{m+1})\right.\nonumber\\
      &&\left.+\cdots+v(S_m+\cdots+S_{m+q-1})\right].
\end{eqnarray}
Taking $f(S;\beta)\equiv f(S)=e^{-\beta\,v(S)}$, we find
\begin{equation}
P_{N}\left(S_1,\cdots,S_N;\beta\right)=\frac{1}{Z_N(L;\beta)}\prod^{N}_{m=1}F(S_m,\cdots,S_{m+q-1}),
\end{equation}
where we define $F(S_m,\cdots,S_{m+q-1})=f(S_m)f(S_m+S_{m+1})\cdots f(S_m+\cdots+S_{m+q-1})$.
The joint probability distribution of $n$ consecutive spacings $P_{n}\left(S_1,\cdots,S_n;\beta\right)$ is given by
\begin{equation}
P_{n}\left(S_1,\cdots,S_n;\beta\right)=\int dS_{n+1}\cdots dS_{N}\,P_{N}\left(S_1,\cdots,S_N;\beta\right).
\end{equation}
By definition, the $n^{th}$ spacing distribution function $\hat{p}^{(n)}(S)\equiv\hat{p}^{(n;q)}(S)$ can be written as
\begin{equation}
\hat{p}^{(n;q)}(S)=\int^{\infty}_{0} dS_1\cdots dS_{n+1}\,\delta\left(\eta\right)P_{n+1}\left(S_1,\cdots,S_{n+1};\beta\right),
\end{equation}
with $\eta=S-\sum^{n+1}_i S_i$. Note that this notation makes explicit the dependence of the spacing distributions on the number of interacting neighbors $q$. The average spacing between particles is $\left\langle S\right\rangle=L/N$. Then, the scaled probability density is
\begin{equation}\label{pnsdef}
p^{(n;q)}(s)=\int^{\infty}_{0} dS_1\cdots dS_{n+1}\, \delta(\lambda)P_{n+1}\left(S_1,\cdots,S_{n+1};\beta\right).
\end{equation}
with $\lambda=\eta/\left\langle S\right\rangle$. Note that Eq.~(\ref{pnsdef}) satisfies the normalization conditions \cite{abul}
\begin{equation}\label{grgen}
\int^{\infty}_{0}ds\,p^{(n;q)}(s)=1 \quad \mathrm{and} \quad  \int^{\infty}_{0}ds\,s\,p^{(n;q)}(s)=n+1.
\end{equation}

In general, the integral given in Eq.~(\ref{pnsdef}) cannot be easily calculated by analytical methods. In fact, it can be solved just in few cases. A simple example of this is given by the potential $v(r)=-\ln(r)$ with $q=1$. By using the Laplace transform method it can be shown \cite{Bogomolny,guy} that
\begin{equation}\label{psp1}
p^{(n;1)}(s;\beta)=\frac{(1+\beta)^{(1+\beta)(1+n)}}{\Gamma[(1+\beta)(1+n)]}s^{\beta+n(1+\beta)}e^{-s(1+\beta)}.
\end{equation}
The case $q=N-1\rightarrow \infty$ was solved by Dyson and so is termed ``\textit{Dyson's Brownian Motion Model}" \cite{dyson1,dyson2}. The general case, $1<q<N-1$, with $N,L\rightarrow \infty$ was solved in Ref.~\cite{Bogomolny} through integral equations. Henceforth, we denote as Dyson's Brownian model a 1D system in which the particles interact through a logarithmic potential, regardless of the range of interaction $q$.

\section{Independent Interval Approximation}
\subsection{General Formalism}
The easiest method to obtain approximate expressions for $p^{(n;q)}(s)$ is the independent interval approximation (IIA). In the IIA, all spacing distributions are generated from the nearest-neighbor distribution $p^{(0;q)}(s)$. The essence of the approach is to neglect the correlations between the sizes of adjacent spacings. Consequently, the probability to find particles around positions $x_1, x_2,\cdots, x_N$ is given by
\begin{eqnarray}\label{IIA1}
P^{\rm eff}_N(x_1, x_2,\cdots, x_N;\beta)&=&\prod^N_{i=1}p^{(0;q)}(x_{i+1}-x_i)\nonumber\\
&=&e^{-\beta\sum^N_{i=1}v_{\rm eff}(x_{i+1}-x_i)}
\end{eqnarray}
Comparing Eq.~(\ref{IIA1}) with the Boltzmann factor given in Eq.~(\ref{joint}), we conclude that
\begin{equation}\label{pottot}
V(x_1,\cdots,x_N)+C=\sum^N_{i=1}v_{\rm eff}(x_{i+1}-x_i).
\end{equation}
The constant $C$ is needed to ensure normalization of the joint probability distribution.

In IIA, each particle is taken to interact with its nearest neighbors via an effective pair potential given by $v_{\rm eff}(r)=-\ln\left(p^{(0;q)}(r)\right)$. This effective potential is generally different from the real potential; most importantly, the effective potential usually depends on the inverse temperature of the studied system. Note that the IIA reproduces exactly the functional form of the interaction potential $V(x_1,\cdots,x_N)$ in the case of $q=1$. However, we caution that entropic repulsions (arising from fermionic non-crossing interactions such as occur in the TSK model of steps discussed in Sec.~\ref{TSK}) intrinsically involve all particles; hence, this part of $V(x_1,\cdots,x_N)$ in Eq.~(\ref{pottot}) cannot be decomposed exactly into components between neighboring parts.

\subsection{Applications}

As a sample application of the IIA, consider Dyson's Brownian model with $q=1$. From Eq.~(\ref{pnsdef}) the statistical behavior of this model is clearly equivalent to that in which particles interact only with their nearest neighbors through an effective pair potential given by
\begin{equation}\label{dysonp1IIA}
v_{\rm eff}(r)=-\ln(r)+\frac{(1+\beta)}{\beta}\,r+C,
\end{equation}
In Eq.~(\ref{dysonp1IIA}), the first term represents the repulsive force between adjacent particles while the second takes into account the effect of the average pressure force on the ends of the interval of length $r$ due to the particles which are outside of the interval. The statistical behavior of Dyson's Brownian model with nearest-neighbor interactions in the IIA is given by
\begin{equation}\label{IIAdyson}
P^{\rm eff}_N(x_1, x_2,\cdots, x_N;\beta)=C^N\prod^{N}_{j=1}e^{\beta\ln(S_{j})-(1+\beta)S_{j}}.
\end{equation}
From this discussion it should be clear that many systems can share the same statistical behavior even when they have different interaction potentials between particles. In the above example, the real and effective system have the same statistical behavior but their interaction potentials differ.

As for all systems with just nearest-neighbor interactions, in the IIA all spacing distribution functions can be calculated analytically in Laplace space \cite{Bogomolny,salsburg}.
This approximation should work properly in systems with short-range interactions between particles because the correlations between the sizes of adjacent spacings are usually weak compared to the case of long- or infinite-range interactions.

\begin{figure}[htp]
\begin{center}
$\begin{array}{c}
\includegraphics[scale=0.28]{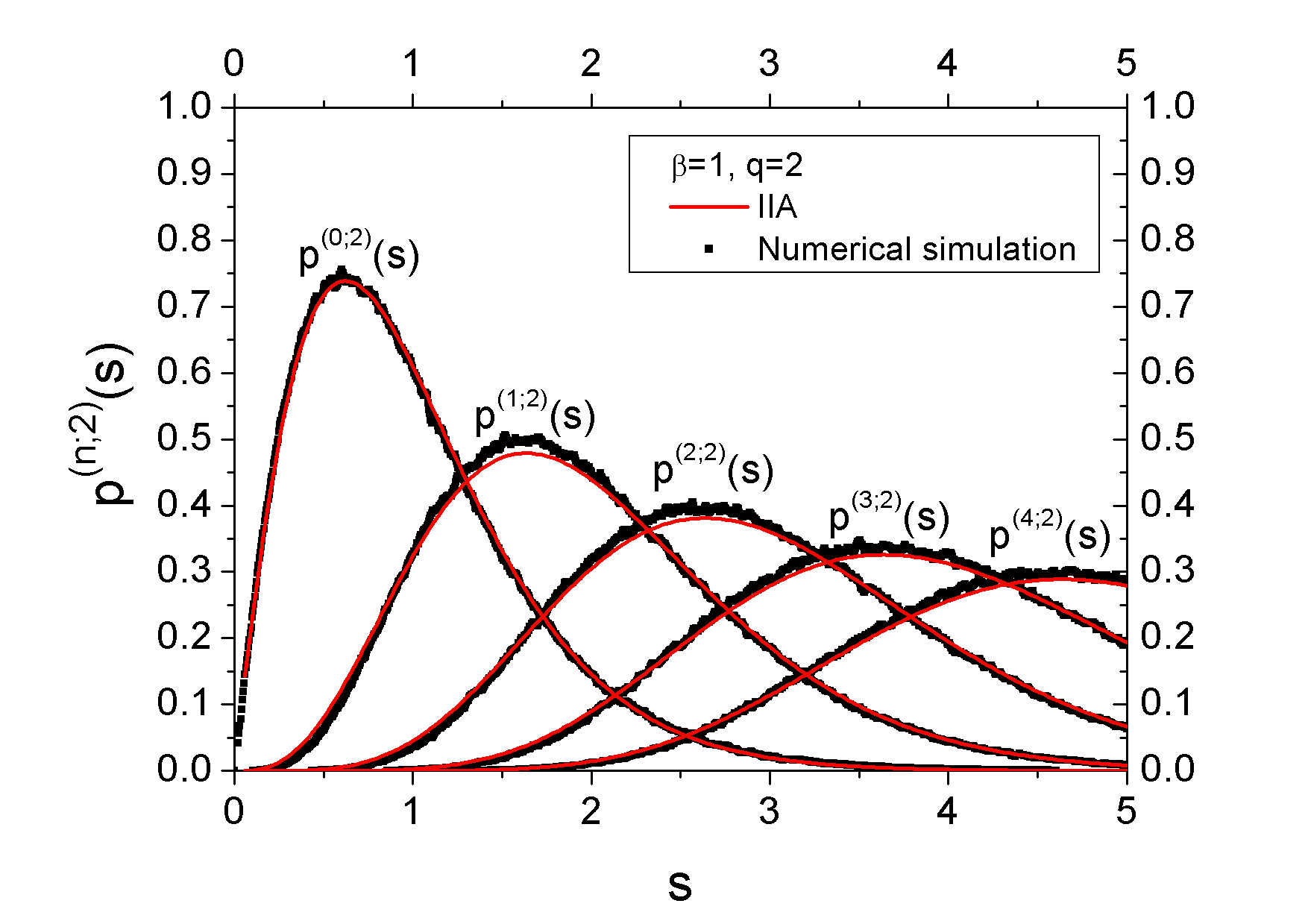}\\
(a)\\
\includegraphics[scale=0.28]{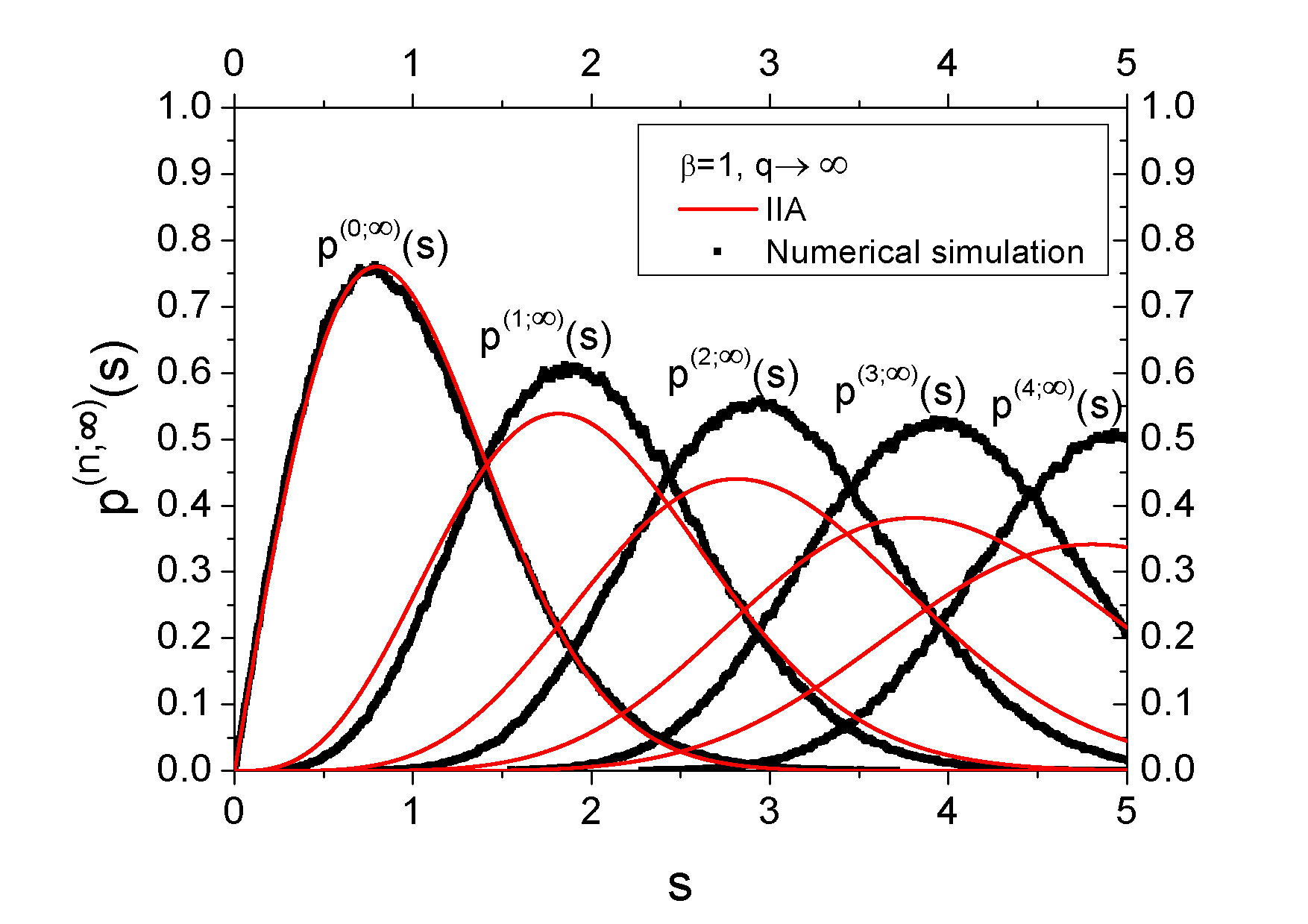}\\
(b) \\
\end{array}$
\end{center}
\caption{(Color online) The IIA for Dyson's Brownian motion model with a) $q=2$  and b) $q=49$.}
\label{modelIIA}
\end{figure}

Consider again Dyson's Brownian motion model. By integrating Eq.~(\ref{IIAdyson}) we see that, for $q=1$, the IIA describes exactly the statistical behavior of this system.  Fig.~\ref{modelIIA}(a) shows that the IIA is also a good approximation for $q=2$. The functions plotted in Fig.~\ref{modelIIA}(a) were calculated from the exact expression for $p^{(0;2)}(s)$ given in Ref.~\cite{Bogomolny} for $q=2$
\begin{equation}\label{p0dys}
p^{(0;2)}(s)=s\, (2.4773+6.0681\,s+3.7159\,s^2)\, e^{-3\,s},
\end{equation}
through Eqs. (\ref{pnsdef}) and (\ref{IIA1}).
In this case, the exact expression for $p^{(1;2)}(s)$ is given by
\begin{equation}\label{p2p1excr}
p^{(1;2)}(s)=s^4\, (2.5054 + 3.068\,s + 0.7516\,s^2)\, e^{-3 s},
\end{equation}
while in the IIA we have as result
\begin{eqnarray}\label{p2p1exc}
p^{(1;2)}(s)&=& s^3\,(1.02247 + 2.50462 s + 2.14728 s^2 + \nonumber\\
 &&0.751391 s^3 + 0.0986 s^4)\,e^{-3\,s}.
\end{eqnarray}
We conclude that the IIA reproduces the $p^{(1;2)}(s)\propto e^{-3\,s}$ behavior of the spacing distribution functions for large values of $s$ but fails for small values of $s$, predicting  $p^{(1;2)}(s)\propto s^3$ rather than the exact result $p^{(1;2)}(s)\propto s^4$.

However, as $q$ increases the IIA becomes very poor, as shown in Fig.~\ref{modelIIA}(b) where we took the widely used Wigner surmise expression for $q=N-1\rightarrow \infty$
\begin{equation}
p^{(0;\infty)}(s)=\frac{\pi}{2}s\,e^{-\frac{\pi}{4}\,s^2},
\end{equation}
and calculating $p^{(n;\infty)}(s)$ from Eqs. (\ref{pnsdef}) and (\ref{IIA1}). As expected from the lack of correlation, the IIA becomes a progressively poor approximation as $n$ increases.

\begin{figure}[htp]
\begin{center}
$\begin{array}{c}
\includegraphics[scale=0.28]{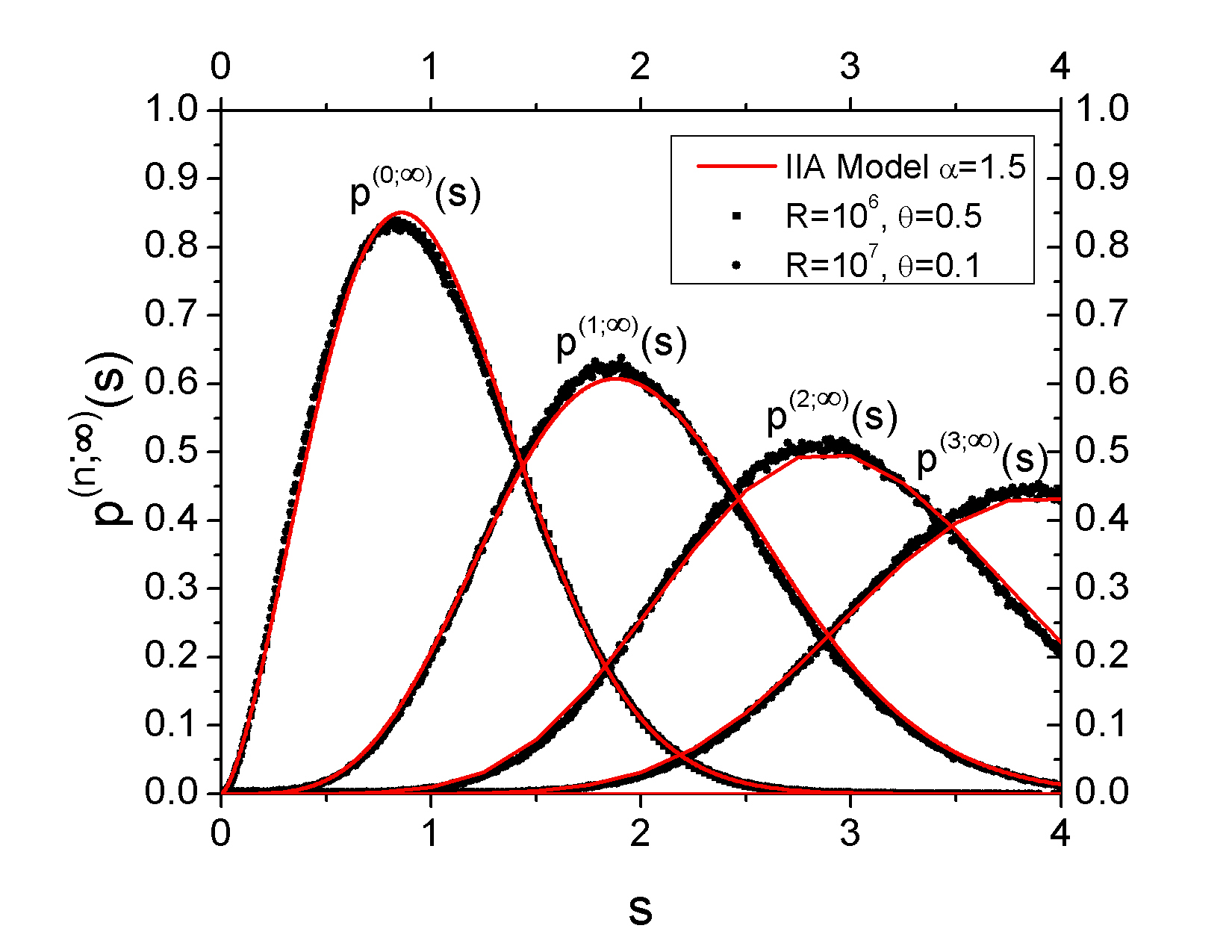}\\
(a)\\
\includegraphics[scale=0.28]{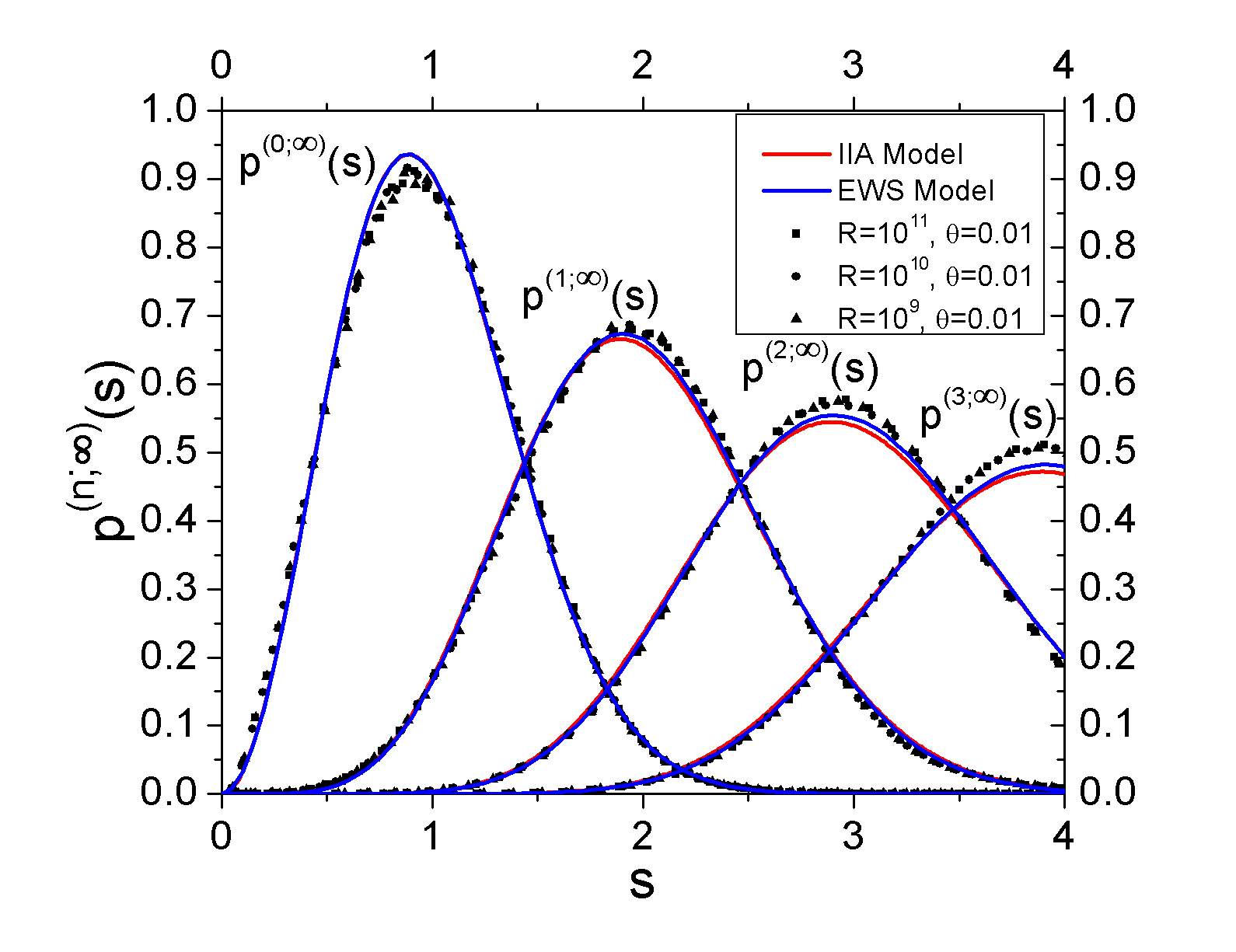}\\
(b) \\
\end{array}$
\end{center}
\caption{(Color online) The IIA for the point island model of epitaxial growth, $d=1$ on the top and $d=2$ on the bottom. We also show the EWS approximation for this model in the 2D case.}
\label{modelIIA2}
\end{figure}

Another example of a system in which the IIA can be applied successfully is the point island model for epitaxial growth \cite{blackman,shi,einstein1,evans1,pimpinelli2}.  In this model, the monomers are deposited at random on a $d$-dimensional lattice. The monomers diffuse across the lattice until reaching a site occupied by another monomer, at which point an island is formed. This coalescing process is called nucleation.  While the monomers are mobile, the islands remain static and do not grow laterally.  Fig.~\ref{modelIIA2} shows that the IIA is a good approximation for this model. In Fig.~\ref{modelIIA2}(a) we use the heuristic equation $p^{(0;\infty)}(s)=A_{\alpha} s^{\alpha}e^{-B_{\alpha}s^2}$ with $\alpha=3/2$ given in Ref.~\cite{einstein1} while in Fig.~\ref{modelIIA2}(b) we use the same expression but with $\alpha=2$ \cite{por}. Additional examples of the IIA can be found in Refs.~\cite{gonzalez,gonzalez3,ben3}.

In short, in the IIA the statistical behavior of an $N$-particle system is modeled by using another system whose particles interact through an effective potential with their nearest neighbors. The effective potential is a function of the inverse temperature and  must be calculated for each particular system. However, as shown in Refs.~\cite{gonzalez,gonzalez3}, $p^{(0;q)}(s)$ does not uniquely define the system.

\section{The Wigner surmise}
The exact expression for the spacing distribution functions for Dyson's Brownian motion model with full-range interactions is complicated and unwieldy. However, Wigner proposed accurate and simple expressions for $p^{(0;\infty)}(s)$ in the special  cases of $\beta=1,2$ and $4$. Abul-Magd and Simbel~\cite{abul} subsequently extended Wigner's surmise to $n>0$ by making the ansatz
\begin{equation}
p^{(n+1;\infty)}(s) \; \mbox{\raisebox{-1.5ex}{$\stackrel{\textstyle{\propto}}{\scriptstyle s\rightarrow 0}$}} \; s^{(n+1)\beta}\int^{s}_{0}p^{(n;\infty)}(s) ,
\end{equation}
using Wigner's expressions for $p^{(0;\infty)}(s)$, and assuming Gaussian decay. In this EWS, $p^{(n;\infty)}(s)$ for Dyson's Brownian model with $q=N-1\rightarrow \infty$ is written as
\begin{equation}\label{pwignersurmise}
p^{(n;\infty)}(s)=A_{n}\,s^{\alpha_n}e^{-B_{n} s^2},
\end{equation}
with $\alpha_{n}\! =\! n\! +\! (n\! +\! 1)(n\! +\! 2)\beta/2$. The constants $A_{n}$ and $B_{n}$ are calculated from the normalization conditions given by Eq.~(\ref{grgen}). In the EWS, the functions $p^{(n;\infty)}(s)$ are completely described by their behavior in the two limits $s\rightarrow0$, where $p^{(n;\infty)}(s)\propto s^{\alpha_n}$, and $s\rightarrow \infty$, where $p^{(n;\infty)}(s)\propto e^{-B_n\,s^2}$.

As mentioned above, the EWS was developed to find analytical expressions for the spacing distributions of Dyson's Brownian motion model. Nonetheless, it can also be applied to a great variety of systems with the suitable choice of $\alpha_n$. One advantage of the EWS compared to alternatives like the IIA lies in its mathematical simplicity. Another is that the IIA is well defined only for 1D systems while the EWS does not make any assumptions about the geometry of the system and it can be applied to more general systems. In Fig.~\ref{modelIIA2}(b) we have applied the IIA simply as a mathematical artifact for a 2D case; in this case it is not possible to define spacings between islands.

Fig.~\ref{modelIIA2}(b) shows that the Wigner surmise is a good approximation for the point island model for epitaxial growth with $d=2$. Additionally, in many cases the EWS captures the essential physics of the system. E.g., for Dyson's Brownian motion model with infinite-range interactions, the EWS for $n=0$ is accurate to better than of the 5\% for $\beta=1$ and even better for $\beta$ = 2 or 4 \cite{haake}. For $n>0$ the fits visually seems to have the same quality as $n=0$ \cite{gonzalez,abul}.
In the IIA all $p^{(n;q)}(s)$ can be calculated in Laplace space, but the inverse of those expressions usually cannot be calculated for large values of $n$.  For the point-island model ($d=2$) in the EWS, we invoke Eq.~(\ref{pwignersurmise}) as the ansatz and calculate $\alpha_n$ from a simple argument: Since we know that the IIA gives a good fit for this model, we can expect that for $s\ll1$
\begin{equation}
p^{(n;\infty)}(s)\approx\int_{0}^{\infty} ds_1 \cdots \int_{0}^{\infty} ds_{n+1}\,\prod_{i=1}^{n+1} p^{(n;\infty)}(s_i)  \propto s^{3\,n+2},
\end{equation}
so that $\alpha_n=3\,n+2$ \cite{por}. In Fig.~\ref{modelIIA2}(b) we see that the differences between the IIA and the EWS are almost imperceptible for the point island model in $d=2$.

Unfortunately, the EWS is not always a good approximation. For example, consider the CRW and Dyson's Brownian model with infinite range of interactions and $\beta=1$. In Ref.~\cite{gonzalez2} it was shown that both systems have Gaussian behavior for large values of $s$ and that $\alpha_n$ is given in Eq.~(\ref{pwignersurmise}), with $\beta=1$, for both systems. However, the EWS gives good results for Dyson's Brownian model but not for the CRW.

Another example is Dyson's Brownian motion with $1<q<N-1$, where $p^{(n;q)}(s)\propto e^{-B_{n}\,s}$ for $s\rightarrow\infty$ \cite{Bogomolny}. It is then tempting to extend our previous approach by writing $p^{(n;q)}(s)$ in the more general form
\begin{equation}\label{broad}
p^{(n;q)}(s)=A_{n}\,s^{\alpha_n}e^{-B_{n} s^\gamma},
\end{equation}
here with $\gamma=1$, while $\alpha_n$ is unchanged from Eq.~(\ref{pwignersurmise}). In fact, for finite-range interactions, the functions $p^{(n;q)}(s)$ have the form
\begin{equation}\label{exacpnd}
p^{(n;q)}(s)=A_{n}s^{\alpha_n}Q_{n}(s)e^{-B_{n}s},
\end{equation}
where $Q_{n}(s)$ is a polynomial whose degree depends on $n$ and $\beta$ \cite{Bogomolny}. Thus, for Dyson's Brownian model with finite-range interactions, the spacing distribution functions cannot be described solely by their behavior in the limits $s\rightarrow0$ and $s\rightarrow \infty$; the behavior of $p^{(n;q)}(s)$ for intermediate values of $s$ is clearly important.

Additional applications of the EWS can be found in Ref.~\cite{gonzalez2}, where the EWS was applied to study the statistical behavior of an $N$ particles system with nearest-neighbor interaction $v(r)=\frac{\pi}{4}r^2-\ln\left(\frac{\pi}{2}r\right)$. In this case the EWS gives excellent results with the ansatz $\gamma=2$ and $\alpha_n=2\,n+1$. This potential was used to model two $d=1$ non-equilibrium systems with domain formation \cite{gonzalez}.

%



\section{Mean-field model}
In Refs.~\cite{pimpinelli,pimpinelli1} a mean-field (MF) model is proposed to obtain an analytical expression for the distribution of terrace widths (spacings between adjacent steps) on vicinal surfaces. Following these ideas, we propose a new approximate description for Dyson's Brownian model with finite-range interactions, that is, for arbitrary values of $q$. This system was solved exactly in Ref.~\cite{Bogomolny}. However, the expressions found there for the spacing distribution function in the range $1<q<N$ are unwieldy because they involve integral equations which increase in difficulty for large values of $q$ and $\beta$. Rather (and following Refs.~\cite{pimpinelli,pimpinelli1}), we propose a `\textit{mean-field}' approximation to obtain simple expressions for this model.

\subsection{Nearest-Neighbor Distribution with Finite Range Interaction}
\subsubsection{Simple mean-field approximation (SMF)}
According to the logarithmic potential proposed by Dyson, the force $F_m\equiv F_m(x_1,\cdots,x_N)$ on the $m^{\rm th}$ particle is given by 
\begin{equation}\label{force}
F_m=\sum^{q}_{j=1}\left(\frac{1}{x_{m+j}-x_m}-\frac{1}{x_{m}-x_{m-j}}\right).
\end{equation}
Consequently, the deterministic damped-oscillator equations for the $m^{\rm th}$ particle are
\begin{equation}\label{langevina}
M\frac{d^2x_m}{dt^2}=-\frac{2}{\Gamma} \frac{dx_m}{dt}+F_m\, .
\end{equation}
where $2/\Gamma$---anticipating the subsequent development---is the friction coefficient.  In the overdamped limit of  strong friction, the system relaxes quickly to the equilibrium, allowing us to take $d^2x_m/dt^2=0$ and rewrite Eq.~(\ref{langevina}) as first-order differential equations
\begin{equation}\label{langevinb}
\frac{dx_m}{dt}=\frac{\Gamma}{2}F_m.
\end{equation}
As in Ref.~\cite{paul}, we then consider the time evolution of the spacings between adjacent particles, $S_m=x_{m+1}-x_m$, adding a stochastic term
to produce the set of Langevin equations
\begin{equation}\label{langevinwa}
\frac{dS_m}{dt}=\frac{\Gamma}{2}\left(F_{m+1}-F_m\right)+\frac{\Gamma}{2}\eta_m(t),
\end{equation}
where the $\eta_m(t)$ introduce Gaussian white noise for each spacing, and
\begin{widetext}
\begin{equation}\label{langevinwb}
F_{m+1}-F_m=\frac{1}{\sum_{i=1}^j S_{m+i}}-\frac{1}{S_m+\sum_{i=1}^{j-1} S_{m-i}}-\frac{1}{S_m+\sum_{i=1}^{j-1} S_{m+i}}+\frac{1}{\sum_{i=1}^j S_{m-i}}\, .
\end{equation}
\end{widetext}
Taking the average $\left\langle \right\rangle$ in Eq.~(\ref{langevinwa}) as discussed in the Appendix \ref{app1}, we have
\begin{equation}\label{langevinwc}
\frac{dS}{dt}=C_1\,\Gamma \sum^{q}_{j=1} \frac{1}{j\left\langle S\right\rangle}-C_2\,\Gamma\sum^{q}_{j=1}\frac{1}{S+(j-1)\left\langle S\right\rangle},
\end{equation}
where the $C_j$'s are constants which can be interpreted as the renormalization of the strength of interaction. The first sum at the right of Eq.~(\ref{langevinwc}) can be calculated easily
\begin{equation}
\frac{dS}{dt}=\frac{C_1\,\Gamma}{\left\langle S\right\rangle} H_p-C_2\,\Gamma\sum^{q}_{j=1}\frac{1}{S+(j-1)\left\langle S\right\rangle},
\end{equation}
where $H_q = \sum_{m=1}^q m^{-1}$ is the generalized harmonic number of order $q$. After the change of variables $\tau=\Gamma\,t/\left\langle S\right\rangle^{2}$ and $s=S/\left\langle S\right\rangle$, it is straightforward to show
\begin{equation}\label{langevins1}
\frac{ds}{d\tau}=C_1\,H_q-C_2\,\sum^{q}_{j=1}\frac{1}{s+(j-1)}.
\end{equation}
Then the effective potential associated with  Eq.~(\ref{langevins1}) is
\begin{equation}
v_{\rm eff}(s)=C_1\,H_q\,s-C_2\,\sum^{q}_{j=1}\mathrm{ln}\left(s+(j-1)\right)+C_3,
\end{equation}
The nearest-neighbor distribution $p^{(0;q)}(s)$ is given by
\begin{equation}\label{p0mf}
p^{(0;q)}(s)=A\,\prod^{q}_{j=1}\left(s+(j-1)\right)^{\beta\,C_2}\,e^{-\beta\,C_1\,H_q\,s}.
\end{equation}
The constant $C_2$ can be calculated by using the fact that for low densities $p^{(0;q)}(s)\approx e^{-\beta v(s)}$, which leads to $C_2=1$. The remaining constants can be calculated by using the normalization conditions. In the SMF model, Eq.~(\ref{p0mf}) has a similar functional form to  Eq.~(\ref{exacpnd}). However, the order of the polynomial factor is not the same in both cases. For example, for the case of $q=2$ and $\beta=1$, the SMF model gives
\begin{equation}
p^{(0;2)}(s)\approx3.30\,s\,(1+s)\,e^{-2.45 s},
\end{equation}
which clearly differs from Eq.~(\ref{p0dys}).

\subsubsection{Improved mean-field approximation (IMF)}\label{imfs}
We expect that for large values of $q$ the SMF approximation begins to fail. The reason lies in the fact that implicit in Eq.~(\ref{langevinwb}) there are more terms which depend on $S_m$. For example, in the case $q=2$ we have such terms as
\begin{equation}
\frac{1}{x_{m+2}-x_{m+1}}-\frac{1}{x_{m+2}-x_{m}}=\frac{S_m}{S_{m+1}(S_m+S_{m+1})},
\end{equation}
which clearly depends on $S_m$. The number of such terms increases with $q$.  Furthermore, for full interactions between all pairs of particles, Eq.~(\ref{p0mf}) does not reproduce the behavior of the Wigner surmise. Consequently, we must formulate an improved mean-field approximation (IMF hereafter). After some algebra, it is possible to write Eq.~(\ref{langevinwa}) as
\begin{eqnarray}\label{langevinwd}
\frac{dS_m}{dt}&=&\frac{\Gamma}{2}\sum^{q-1}_{j=1}\frac{S_m}{(x_{m+j+1}-x_{m+1})(x_{m+j+1}-x_m)}\nonumber\\
&+&\frac{\Gamma}{2}\sum^{q-1}_{j=1}\frac{S_m}{(x_m-x_{m-j})(x_{m+1}-x_{m-j})}\nonumber\\
&+&\frac{\Gamma}{2}\left(\frac{1}{x_{m+q+1}-x_{m+1}}+\frac{1}{x_{m}-x_{m-q}}\right)\nonumber\\
&-&\frac{\Gamma}{x_{m+1}-x_{m}}+\frac{\Gamma}{2}\eta_m(t).
\end{eqnarray}
Taking the average $\left\langle \right\rangle$ in Eq.~(\ref{langevinwd}) as discussed in the Appendix \ref{app1}, we find
\begin{equation}
\frac{dS}{dt}=\Gamma\sum^{q-1}_{j=1}\frac{\tilde{C}_1\,S}{j(j+1)\left\langle S^2\right\rangle}+\frac{\Gamma\,\tilde{C}_1}{q\left\langle S\right\rangle}-\frac{\Gamma\,\tilde{C}_2}{S}.
\end{equation}
where the constants $\tilde{C}_1$ and $\tilde{C}_2$ are defined implicitly in Appendix \ref{app1}. After the changes of variables $\tau=\Gamma \,t/\left\langle S\right\rangle^2$ and $s=S/\left\langle S\right\rangle$, then using $\sum^{q-1}_{j=1}[j(j+1)]^{-1}=(q-1)/q$, it is straightforward to show
\begin{equation}\label{langevins}
\frac{ds}{dt}=\tilde{C}_1\left(\frac{q-1}{q}\right)\frac{s}{\left\langle s^2\right\rangle}+\frac{\tilde{C}_1}{q}-\frac{\tilde{C}_2}{s}+\eta.
\end{equation}
Then, the effective potential associated with Eq.~(\ref{langevins}) is
\begin{equation}
v_{\rm eff}(s)=-\tilde{C}_2\,\mathrm{ln}(s)+\frac{\tilde{C}_1}{2}\left(\frac{q-1}{q}\right)\frac{s^2}{\left\langle s^2\right\rangle}+\frac{\tilde{C}_1}{q}s+C.
\end{equation}
Again, using $p^{(0;q)}(s)\approx e^{-\beta v(s)}$ it is easy to find $\tilde{C}_2=1$. The nearest-neighbor distribution $p^{(0;q)}(s)$ is given by
\begin{equation}\label{p1}
p^{(0;q)}(s)=A\,s^{\beta}\exp\left[-\beta\frac{\tilde{C}_1}{2}\left(\frac{q-1}{q}\right)\frac{s^2}{\left\langle s^2\right\rangle}-\beta\frac{\tilde{C}_1}{q}s\right].
\end{equation}
In the case of full-range interactions, the nearest-neighbor distribution can be written as
\begin{equation}\label{p2}
p^{(0;q)}(s)=A\,s^{\beta}e^{-(\beta\,\tilde{C}_1/2)(s^2/\langle s^2\rangle)} \; .
\end{equation}
By using the normalization conditions, one can show that Eq.~(\ref{p2}) is equivalent to the Wigner surmise described by Eq.~(\ref{pwignersurmise}) for $n=0$.  In the case of nearest-neighbor interactions, Eq.~(\ref{p1}) takes the form of Eq.~(\ref{psp1}). For $q>1$, Eq.~(\ref{p1}) predicts a Gaussian tail for large values of $s$ instead of the exponential in Eq.~(\ref{exacpnd}). Despite this discrepancy with the exact result, it is reasonable to expect that the IMF model given by Eq.~(\ref{p1}) leads to good results, especially for large and small values of $q$ (where it reduces to the Wigner surmise and to Eq.~(\ref{psp1}), respectively).

Before generalizing our mean-field models to $n>0$, we offer some comments about Eq.~(\ref{p1}). First, as mentioned before, the IMF for the case of full-range interactions was used previously \cite{pimpinelli,pimpinelli1} to describe the terrace-width distribution between steps on vicinal surfaces. However, Eq.~(\ref{p1}) corresponds to a generalization for the case of finite-range interactions.

Second, the capture zone (CZ) distribution of islands generated in the early stages of epitaxial growth was modeled excellently by using the functional form given by Eq.~(\ref{p1}) \cite{gonzalez7}. There, the authors use the maximum entropy method to justify this functional form as an approximation to the CZ distribution.

Finally, Eq.~(\ref{p1}) has the same functional form as Eq.~(4.1) in Ref.~\cite{izrailev}, where this expression is proposed for Dyson's Brownian model with $0 < \beta\leq 4$ based on heuristic arguments and on the asymptotic form found by Dyson for $p^{(0;\infty)}(s)$ \cite{dyson2}. They found that Eq.~(\ref{p1}) gives good results for Dyson's Brownian model with complete range of interactions for the values of $\beta$ mentioned previously \cite{izrailev}. From our results, this asymptotic form can clearly be interpreted as a mean-field approximation for Dyson's Brownian motion for large values of $q$.

We emphasize that Eq.~(\ref{p1}) is a generalization of the Wigner surmise for finite-range interactions, which seems to be related to many different models.



\renewcommand{\textfraction}{0.05}

\subsection{Arbitrary Spacing Distribution Functions}
\begin{figure*}[!htp]
\begin{center}
$\begin{array}{ccc}
\includegraphics[scale=0.22]{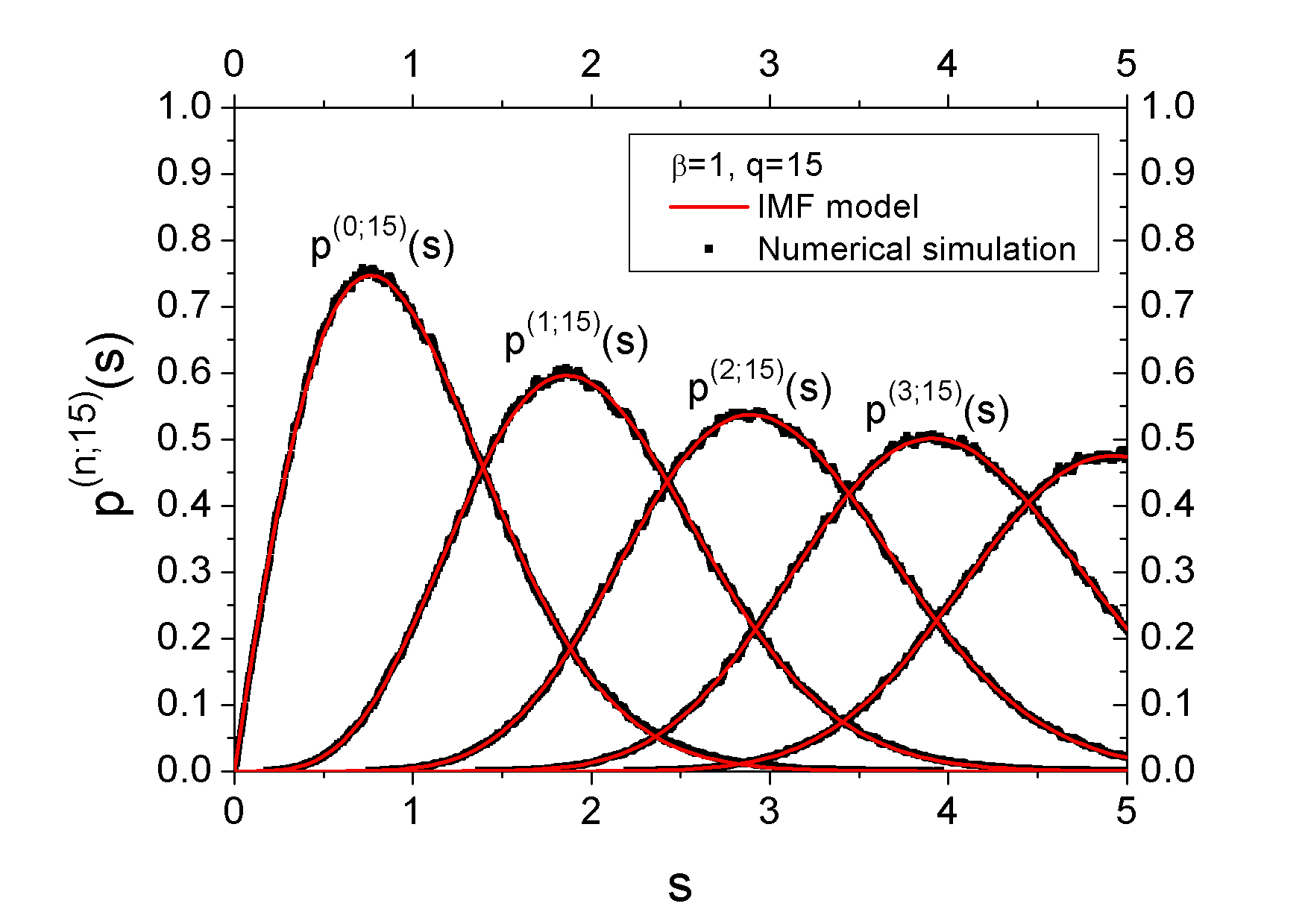}&
\includegraphics[scale=0.22]{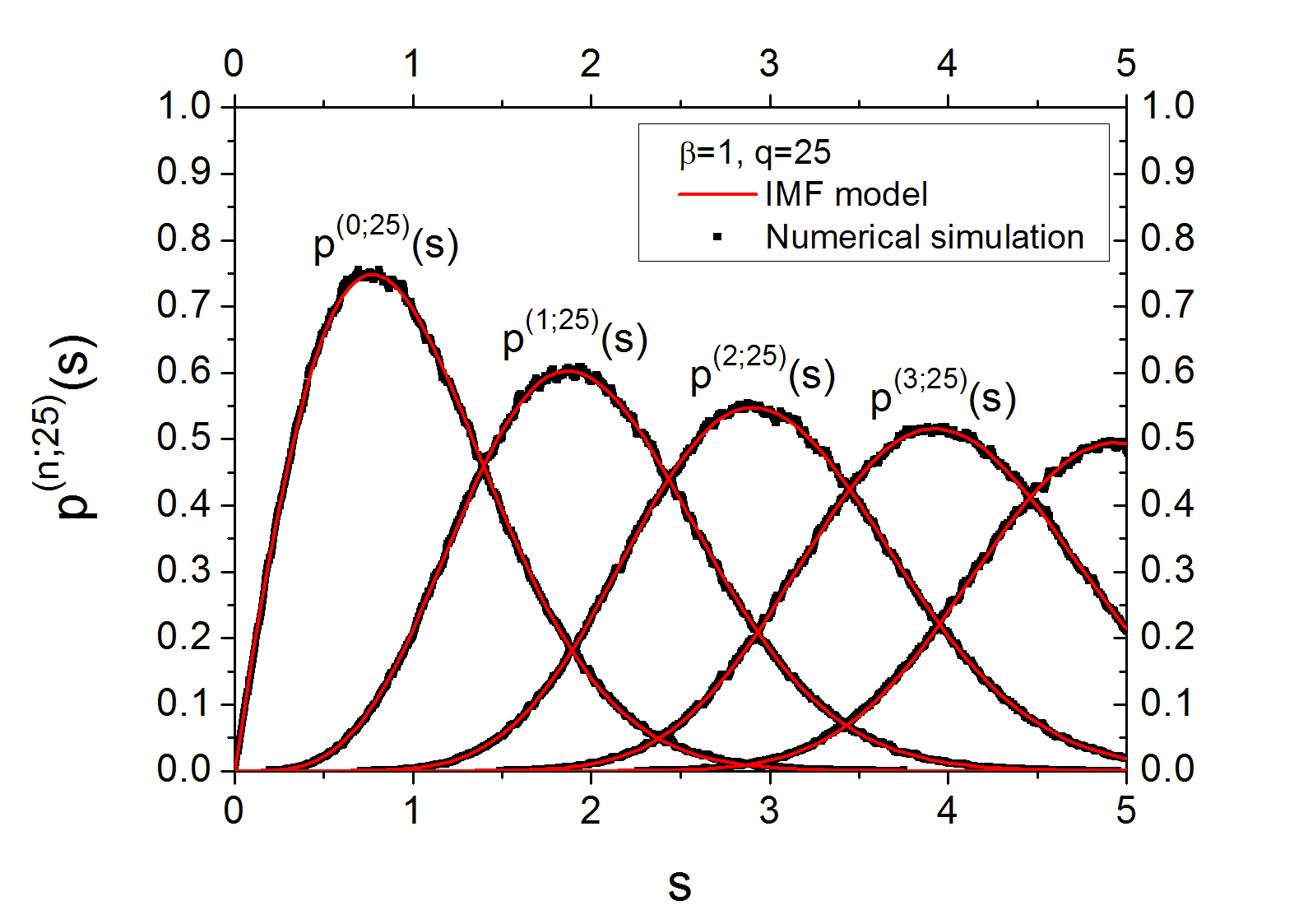}\\
(d) & (e) \\
\end{array}$
\end{center}
\caption{(Color online) Comparison of SMF and IMF models with simulated data, for $\beta$=1, for interactions spanning the range $q$ from nearest-neighbors to essentially infinite, viz.\ values of $q$ of (a) 2, (b) 3, (c) 7, (d) 15, (e) 25.}
\label{MFb1}
\end{figure*}

\begin{figure*}[!htp]
\begin{center}
$\begin{array}{ccc}
\includegraphics[scale=0.22]{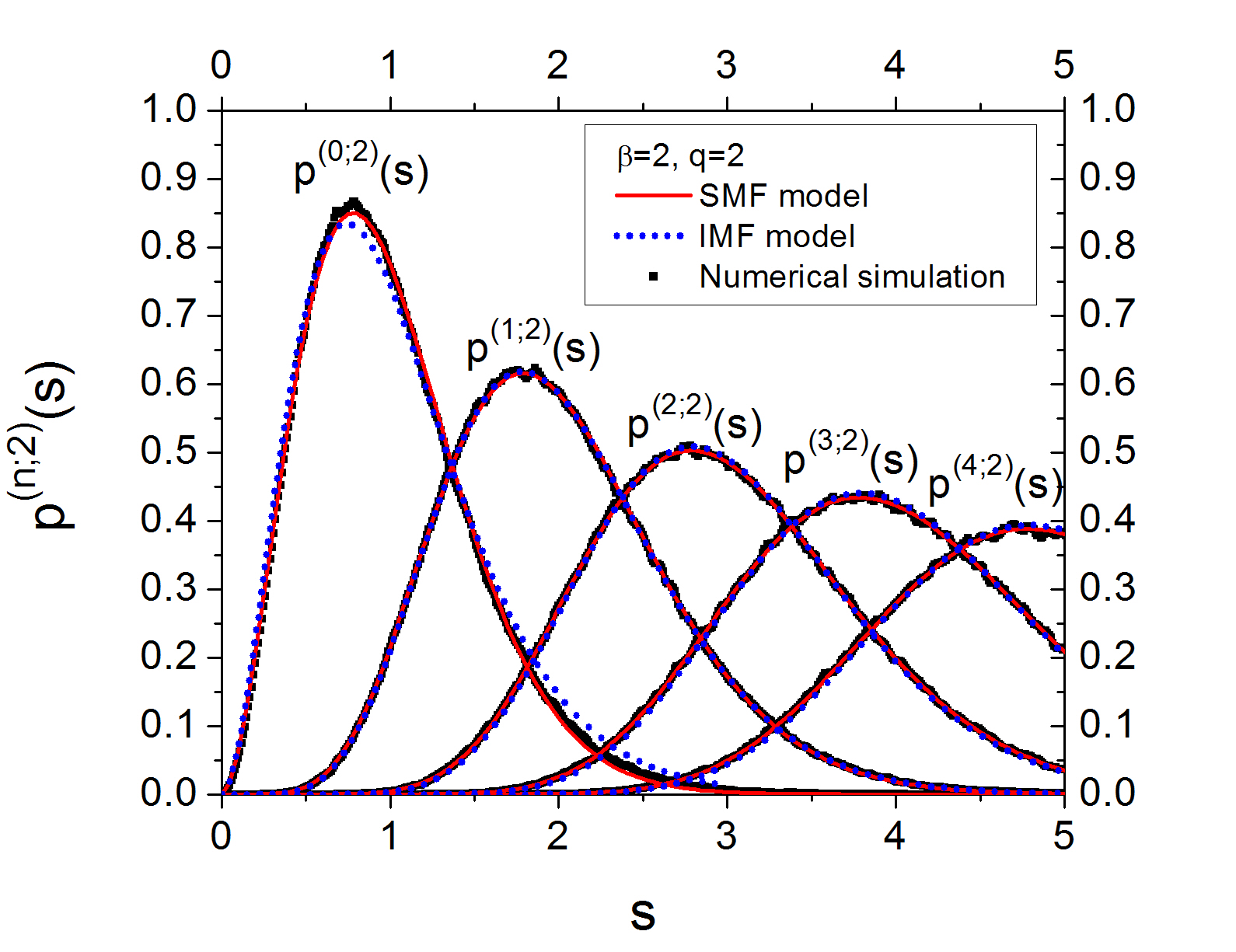}&
\includegraphics[scale=0.22]{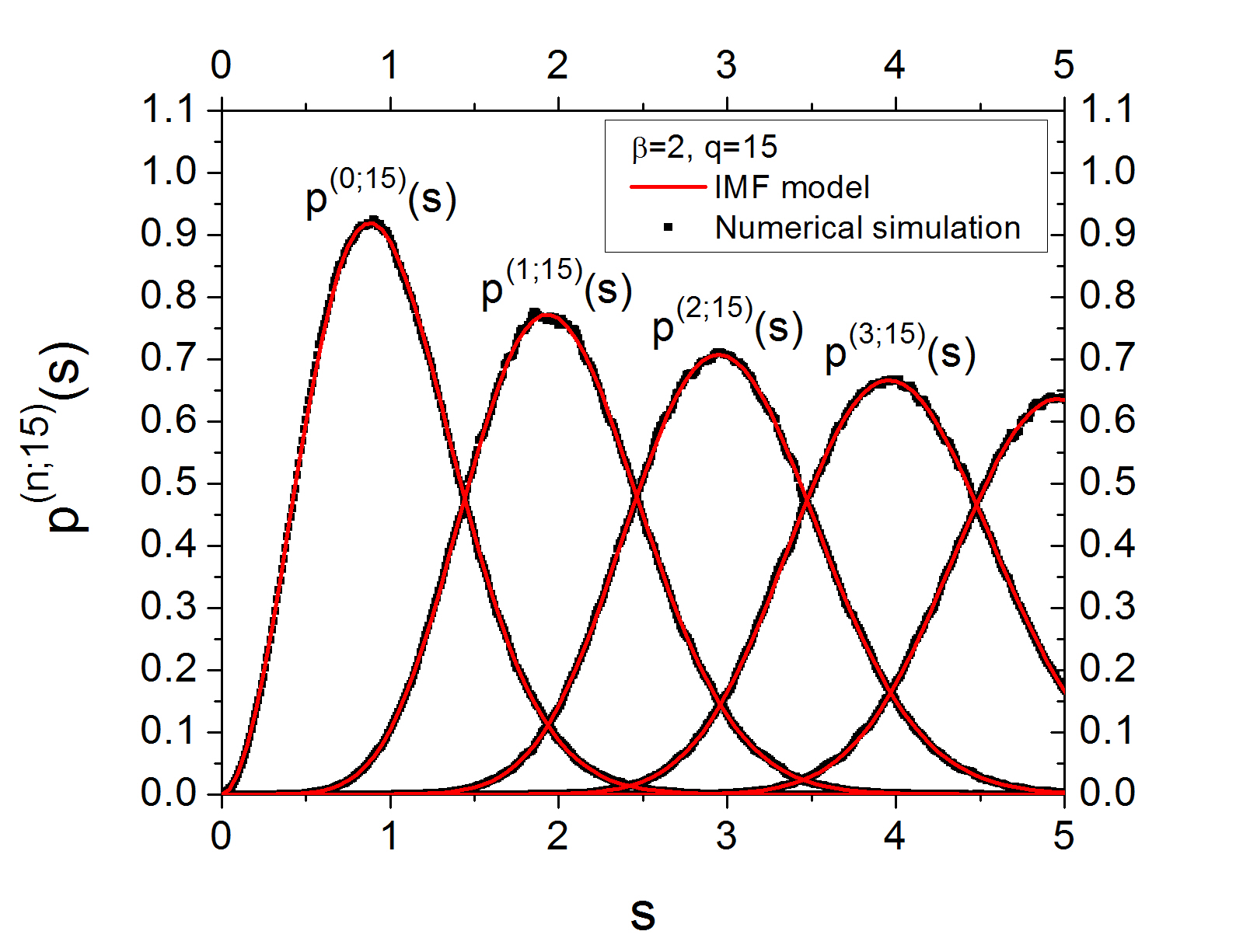}&
\includegraphics[scale=0.22]{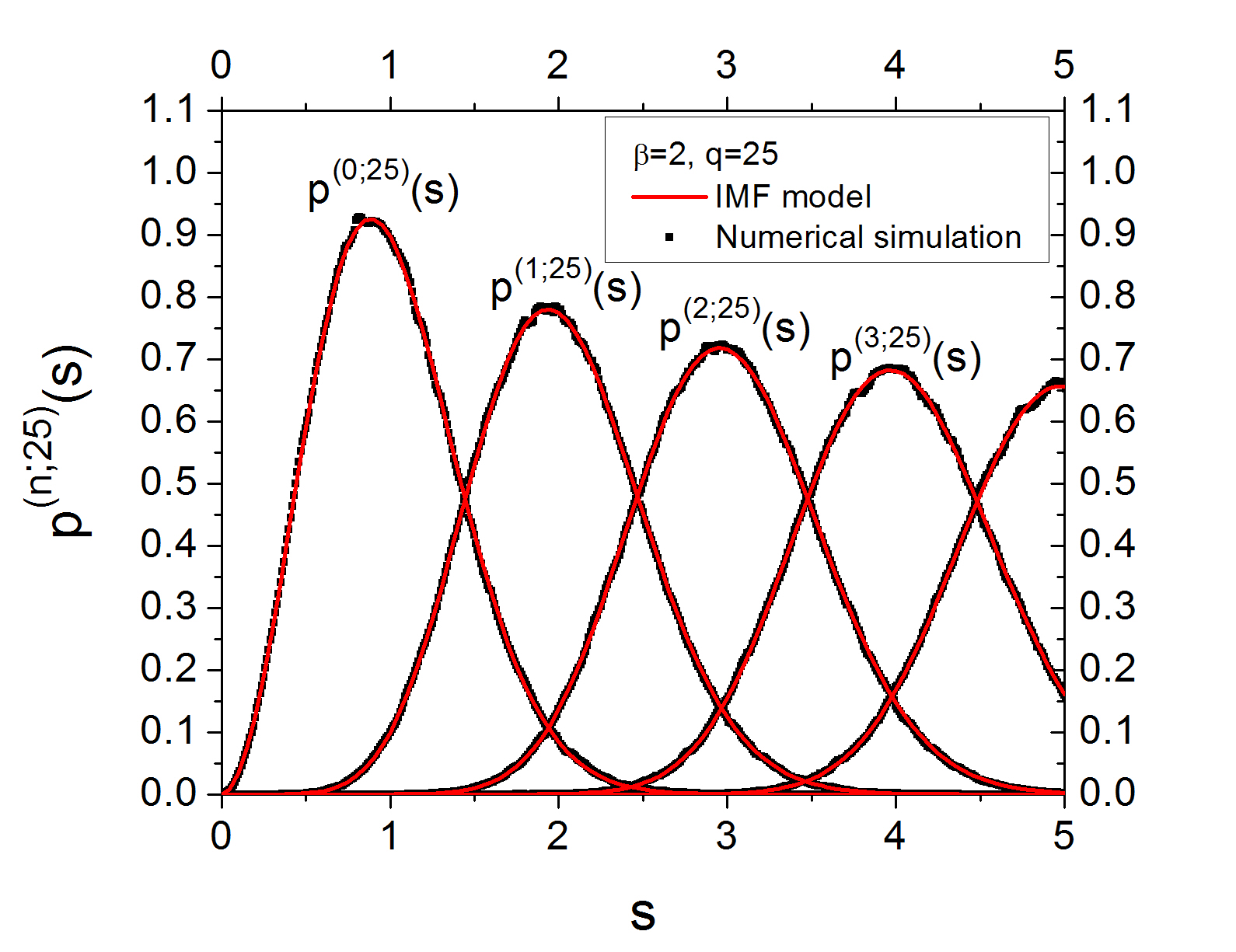}\\
(a) & (b) & (c) \\
\end{array}$
\end{center}
\caption{(Color online) SMF and IMF models, with $\beta$=2, for ranges of interaction $q = 2, 15, 25$.}
\label{MFb2}
\end{figure*}

\begin{figure*}[!htp]
\begin{center}
$\begin{array}{cc}
\includegraphics[scale=0.22]{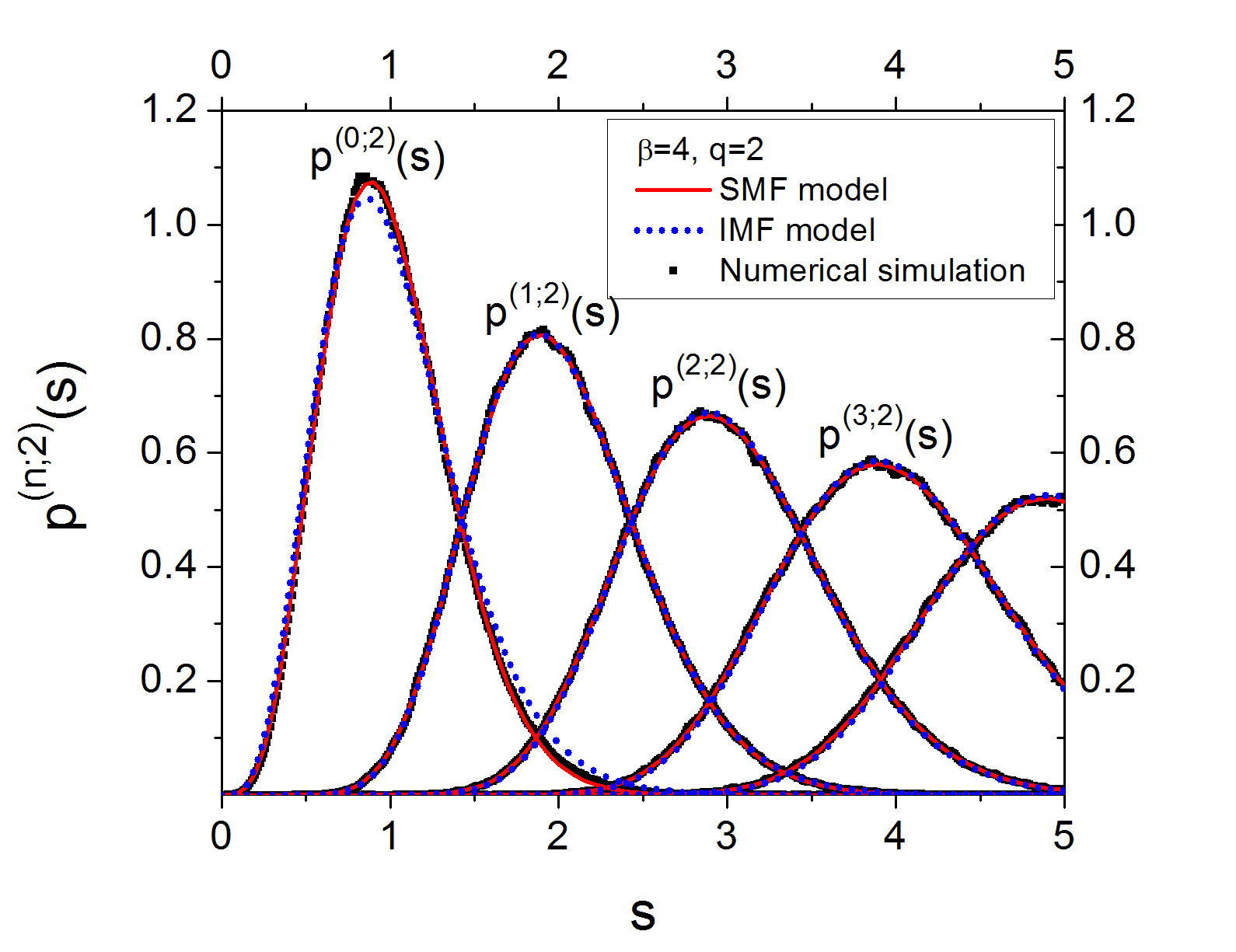}&
\includegraphics[scale=0.22]{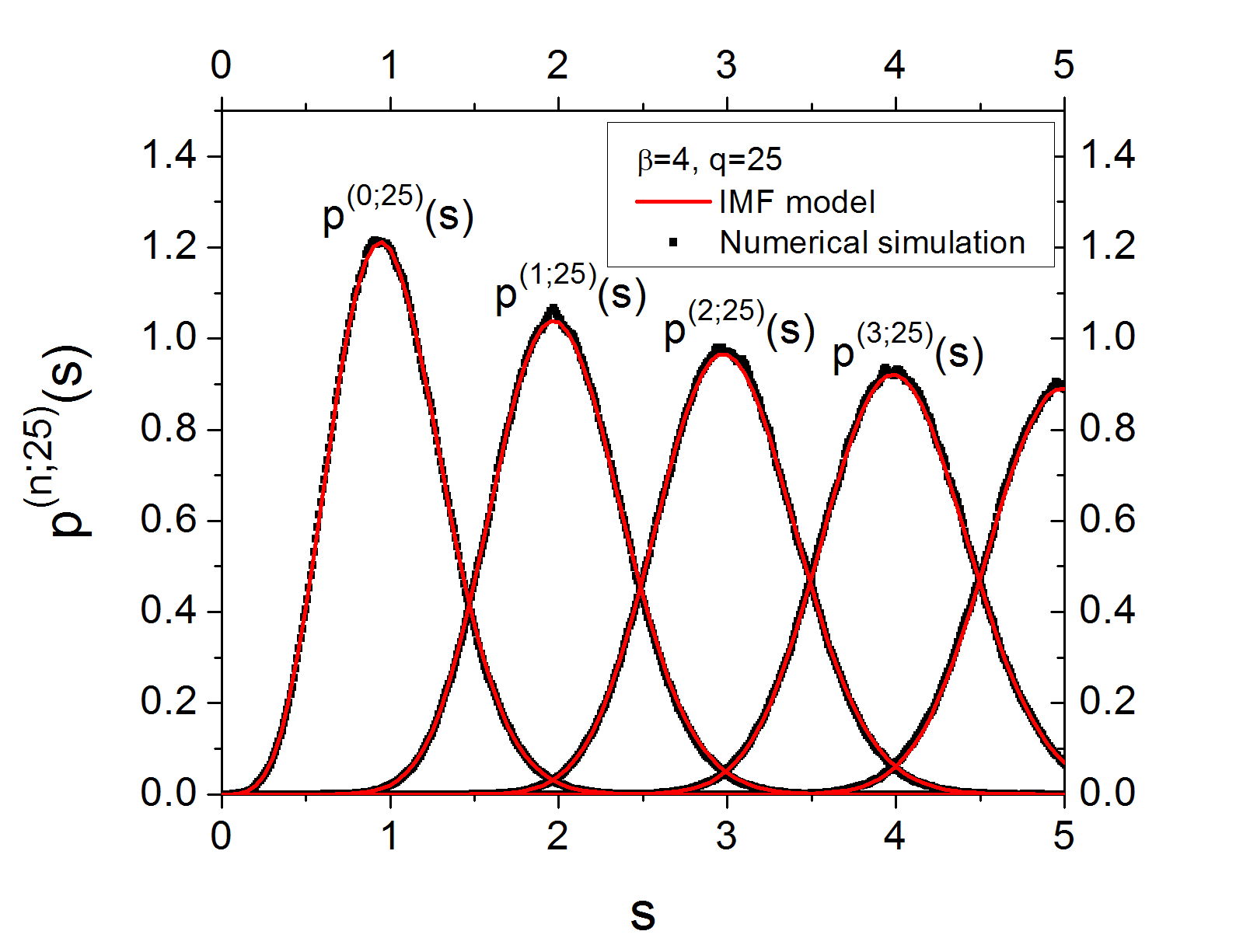}\\
(a) & (b)  \\
\end{array}$
\end{center}
\caption{(Color online) SMF and IMF models for short and long ranges of interaction, with $\beta$=4.}
\label{MFb4}
\end{figure*}

For arbitrary interaction range, the analytic expression for the $n^{th}$ spacing distribution function is not so clear. The spacing distribution functions in the limit $s\rightarrow 0$ have the form
\begin{equation}
p^{(n;q)}(s)\propto s^{\alpha_n},
\end{equation}
where $\alpha_n$ is a function of $\beta$, $n$ and $q$. From Eq.~(\ref{pnsdef}) it is possible to find this exponent by following the same kind of arguments used in Ref.~\cite{gonzalez2} for the CRW case, see Appendix \ref{app2}. However, for Dyson's Brownian motion model it is necessary to distinguish between the two cases $n<q-1$ and $n\geq q-1$. For the first we find
\begin{equation}\label{alphand1}
\alpha_n=\frac{\beta\,(n+1)(n+2)}{2}+n,
\end{equation}
which is the same exponent as for random-matrix ensembles and Dyson's Brownian model with $q\rightarrow\infty$ and $\beta=$1, 2, or 4. However, for $n\geq q-1$ we find
\begin{equation}\label{alphand2}
\alpha_n=\frac{\beta\,q\,(3+2n-q)}{2}+n,
\end{equation}
\noindent differing from its counterpart in random matrices. Hence, the SMF model for arbitrary values of $n$ can be written as
\begin{equation}\label{pnmf}
p^{(n;q)}(s)=A\,s^{\alpha_n}e^{-\beta\,C^n_1\,H_q\,s}\,\prod^{q-1}_{j=1}\left(s+j\right)^{\beta}
\end{equation}
while the IMF model is given by
\begin{equation}\label{pnmfb}
p^{(n;q)}(s)=A_n\,s^{\alpha_n}e^{-\beta\frac{\tilde{C}^n_1}{2}\left(\frac{q-1}{q}\right)\frac{s^2}{\left\langle   s^2\right\rangle_{n}}-\beta\frac{\tilde{C}^n_1}{q}s}
\end{equation}

\noindent where $\alpha_n$ is defined by Eqs. (\ref{alphand1}) and (\ref{alphand2}).
The results of the SMF and IMF models for $\beta=$ 1, 2 and 4 are shown in Figs.~\ref{MFb1},~\ref{MFb2} and \ref{MFb4}, respectively. In general, the IMF model gives better results than the SMF model. However, for large $n$ and small $q$ both give similar results (see Figs.~\ref{MFb2}(a) and \ref{MFb4}). The SMF model describes the spacing distribution functions reasonably well for small values of $q$ and $\beta$ (see Figs.~\ref{MFb1}(a), \ref{MFb1}(b) and \ref{MFb2}(a)). However, as expected from discussion at the beginning in Section \ref{imfs}, the SMF model gives poor results for higher values of $q$ and $\beta$, as shown in Fig.~\ref{MFb1}(c). In general, the IMF gives good results. For large values of $q$, the IMF model gives excellent results, even for large values of $n$, at least for $\beta=$ 1, 2 and 4. However, for small values of $q$ and $\beta=2, 4$, Figs.~\ref{MFb2}(a) and Figs.~\ref{MFb4}(a) show that it gives good results only for small $n$. The SMF and IMF models give the exact statistical behavior of the system in the case $q=1$ while for full-range interactions, only the IMF reduces to the EWS.

\renewcommand{\textfraction}{0.00}
\renewcommand{\topfraction}{0.99}
\setcounter{totalnumber}{5}

\subsection{Another Example: TSK model}\label{TSK}
Vicinal crystals usually have terraces oriented in the high-symmetry direction separated by steps of atomic height, see for example Ref.~\cite{williams}.   The terrace-width distribution (TWD) of vicinal surfaces, that is, the distribution of separations between adjacent steps has special interest for experimentalist because it gives information about the the interaction between steps \cite{einstein}.  It is assumed that the steps cannot cross each other and, in the simplest case of the terrace-step-kink (TSK) model, that the dominant excitation is the formation of kinks.  Typically the step edges interact via a pair potential $v(x)$, where $x$ is the distance between the steps. Usually, the

\begin{figure*}[!htp]
\begin{center}
$\begin{array}{ccc}
\includegraphics[scale=0.2]{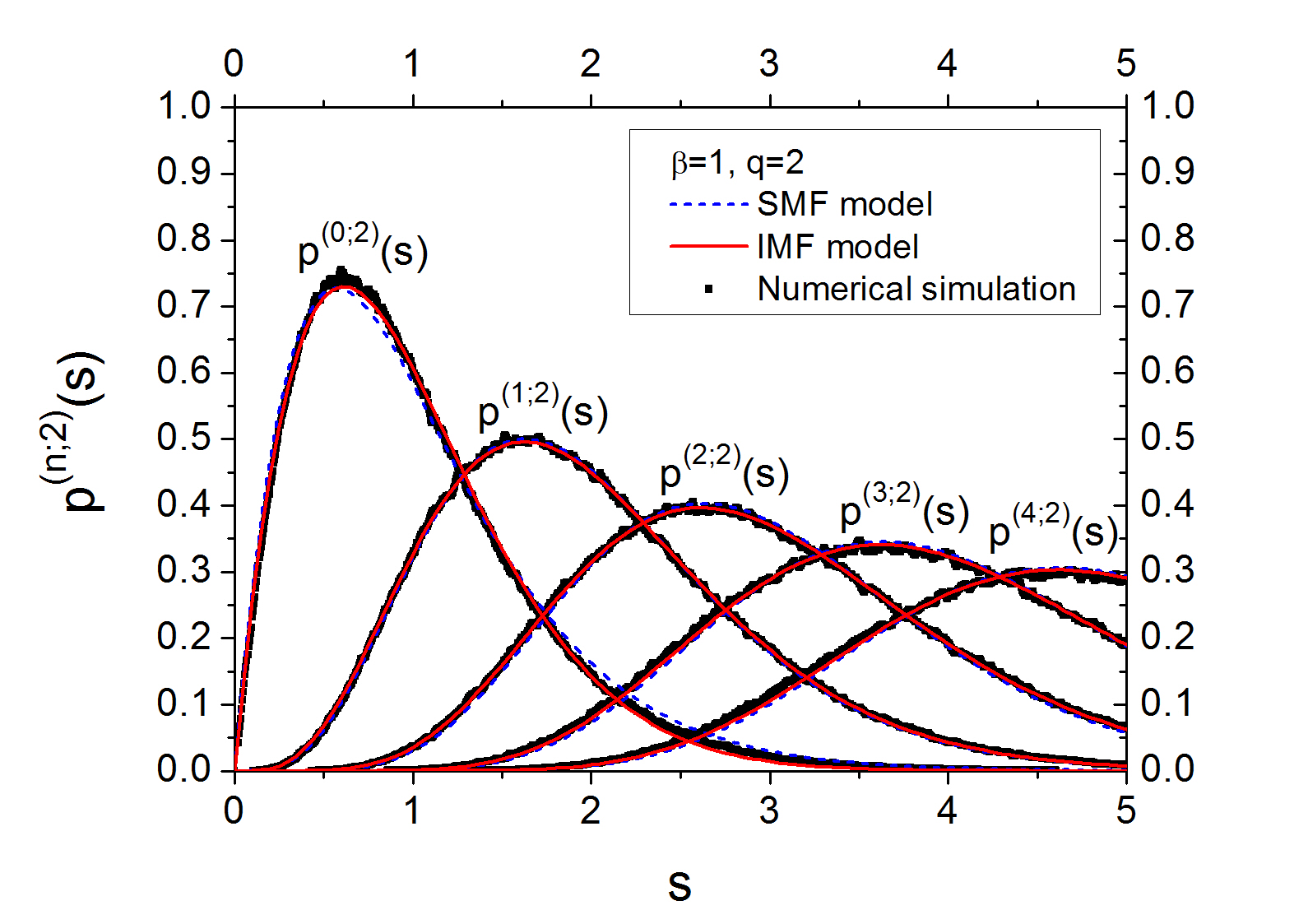}&
\includegraphics[scale=0.2]{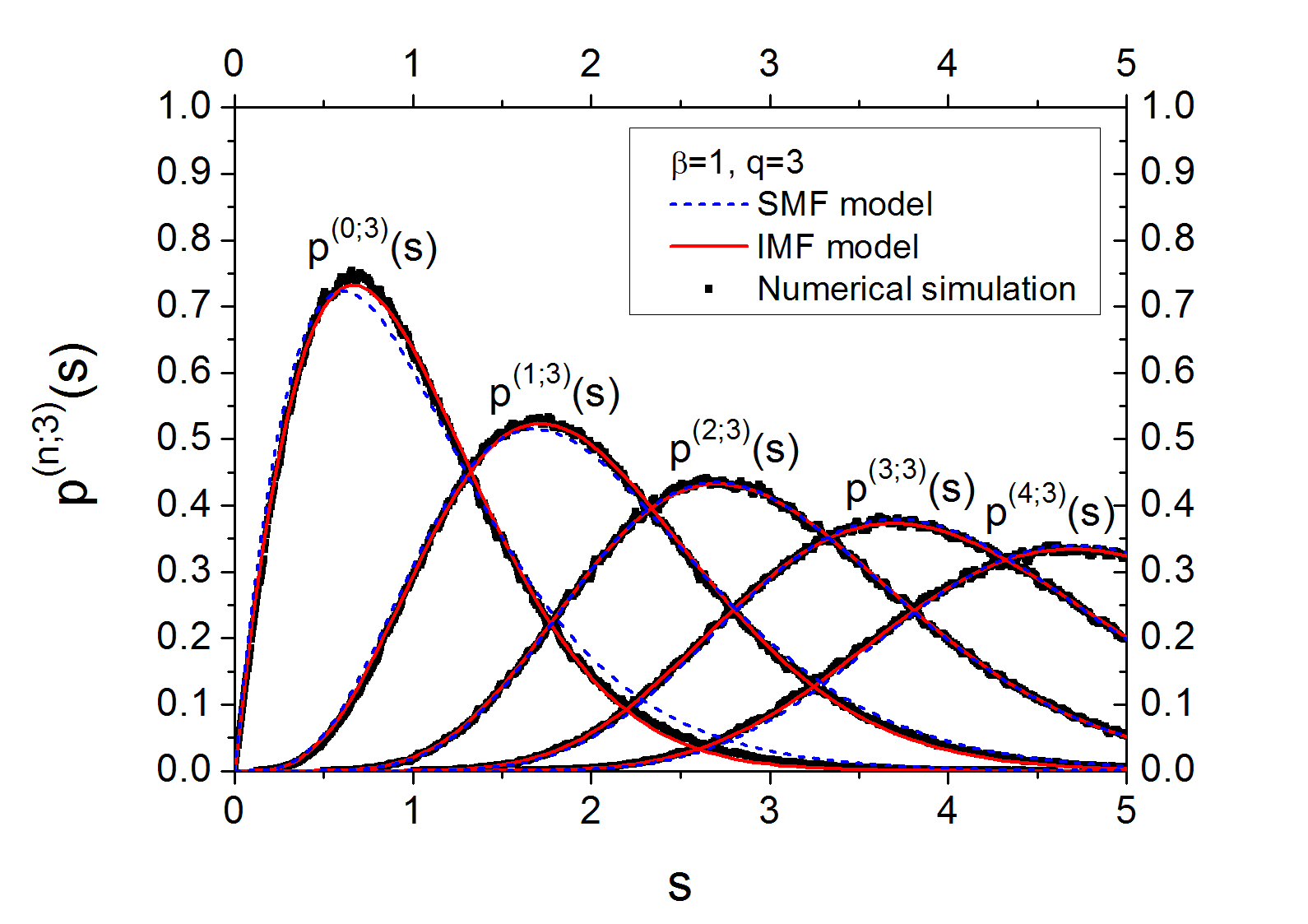}&
\includegraphics[scale=0.2]{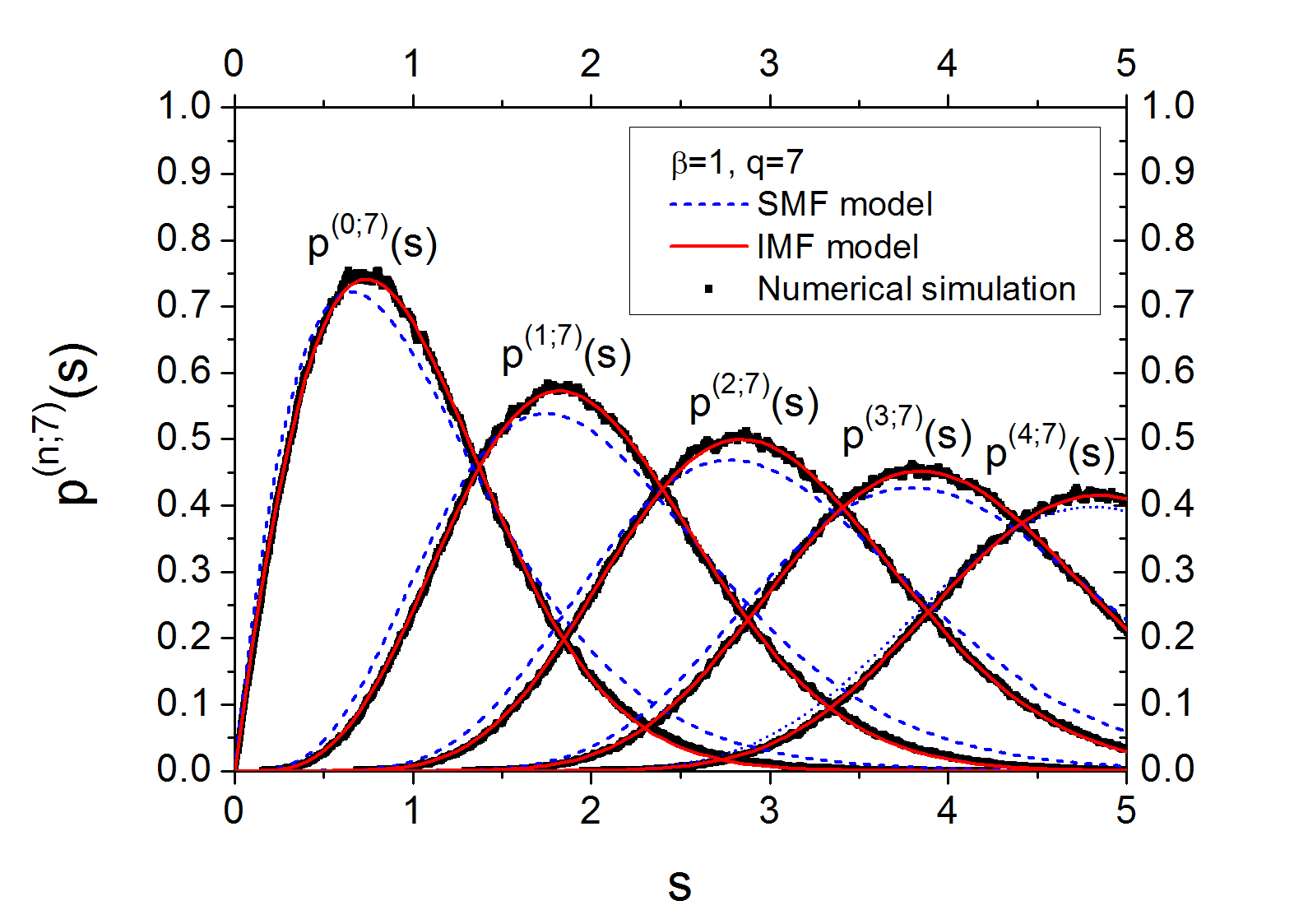}\\
(a) & (b) & (c) \\
\end{array}$
\end{center}
\end{figure*}
\vspace{-10mm}
\pagebreak

\begin{figure}[hbp]
\begin{center}
\includegraphics[scale=0.22]{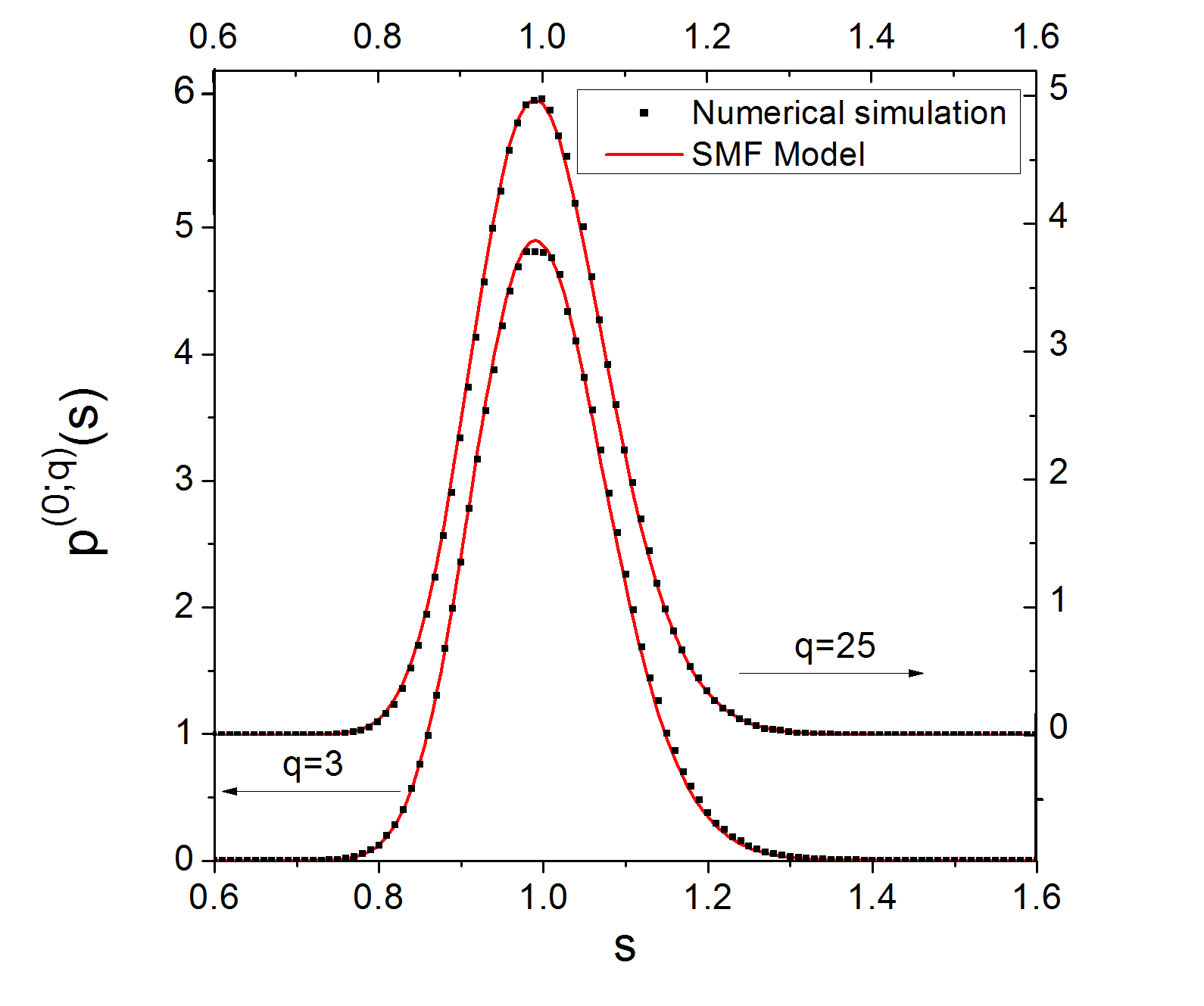}\\
\end{center}
\caption{(Color online) The SMF model for a set of particles which interact via $v(s)\propto s^{-2}$, with $q=3$ and, offset upward one unit, $q=25$. In both cases we used $\beta=1$ in the numerical simulation.}
\label{model1}
\end{figure}

\noindent elastic and entropic contributions to the step-step interactions take the form $v(x)=\mathcal{A}\,x^{-2}$. In Refs.~\cite{paul,yancey}, it was shown that this kind of potential gives a good description of the TWD for interacting steps. That is our motivation to extend our SMF model to this potential. We used the same approximation made in the previous section but now with this potential, finding

\begin{equation}
p^{(0;q)}(s)= A\,e^{-\beta\,\mathcal{A}\,C_3\,H_q^{3}\,s- (\beta/2)\,\mathcal{A}\,\sum^{q}_{j=1}\left(s+(j-1)\right)^{-2}}.
\label{smlp}
\end{equation}
The functional form of the SMF model is similar to the one of the zeroth-order model proposed in Ref.~\cite{paul}.
Fig.~\ref{model1}(b) shows that Eq.~(\ref{smlp}) accounts well for the numerical data. Remarkably, in spite of its simplicity, the SMF approximation describes $p^{(0;q)}(s)$ very well even for large $q$.

\section{Conclusion}
In many cases our knowledge about the statistical behavior of a many-particle system is reduced to the spacing distribution functions due to our inability to obtain numerically, experimentally, or analytically the joint distribution function $P_N(x_1,\cdots,x_N;\beta)$. Since the information contained in $p^{(0;q)}(s)$ is very limited, $p^{(n;q)}(s)$ with $n>0$ should also be calculated to obtain additional physical information about the system.

In the IIA, the statistical behavior of the system is completely determined by $p^{(0;q)}(s)$. This is a crude approximation which becomes exact only in the case of nearest-neighbor interactions. However,  we showed that the IIA can be applied satisfactorily in many cases. One of these cases is Dyson's Brownian model for low values of $q$. In this case we expect that the correlations between gap sizes are weak, justifying the use of the IIA.  Naturally, for large values of $q$ the IIA gives poor results. We found also that the IIA---with the suitable choice of $p^{(0;q)}(s)$---is a reasonable approximation for $p^{(n;q)}(s)$ in the point-island model for epitaxial growth, at least in the 1D and 2D.

The EWS can be applied in systems where $p^{(n;q)}(s)$ can be characterized by its behavior in the limits $s\ll1$ and $s\gg1$. This is the case of Dyson's Brownian model with $q=1$ and $q=N-1\rightarrow \infty$. However the EWS does not give good results for this model in the case of $1<q<N-1$. The EWS can also be used satisfactorily in the case of the 2D epitaxial growth mentioned previously.

The mean-field approximation proposed in this paper is a general tool to study classical interacting particles with finite or full interaction range. The SMF and IMF models are more general approximations than the IIA or the EWS. In particular, the IMF model given by Eq.~(\ref{pnmfb}) can be interpreted as a generalization of the EWS for finite range of interaction. Additionally, the functional form of Eq.~(\ref{p1}) can be used in different contexts \cite{gonzalez7,izrailev}, suggesting that this kind of model is related with other models for different systems.

In order to generalize the SMF and IMF models for $n>0$ in Dyson's Brownian model, we calculated explicitly $\alpha_n$ from the results given in Ref.~\cite{Bogomolny}. We found that one of the effects of the finite range of interactions is to change $\alpha_n$. For $n<q-1$, $\alpha_n$, has the same functional form of the case of complete range of interactions and it does not depend on $q$. For the case $n\geq q-1$ we found a $q$ dependence in $\alpha_n$.

The IMF gives a good description of the Dyson's Brownian model at least for $\beta=1,2$ and 4. In spite of their simplicity, the SMF and IMF models give excellent results for different kinds of interaction potentials. In particular, we also tested the SMF approximation for the $v(r)=\mathcal{A}/s^2$ potential of interacting steps.



\section{Acknowledgments}
This work was supported by the NSF-MRSEC at the University of Maryland, Grant No.\ DMR 05-20471 and DOE via University of Tennessee, with ancillary support from the Center for Nanophysics and Advanced Materials (CNAM).
\appendix
\section{Mean-field approximation}\label{app1}
As a first approximation we replace $S_{n}\rightarrow\left\langle S\right\rangle=L/N$ for $n\neq m$. In this spirit, the average value over the ensemble of particles at time $t$ of the quotients in Eq.~(\ref{langevinwb}) can be written as
\begin{equation}
\left\langle \frac{1}{\sum_{i=1}^j S_{m\pm i}}\right\rangle\rightarrow\frac{C_1}{j \left\langle S\right\rangle},
\end{equation}
and
\begin{equation}
\left\langle\frac{1}{S_m+\sum_{i=1}^{j-1} S_{m\pm i}}\right\rangle\rightarrow\frac{C_2}{S+(j-1) \left\langle S\right\rangle},
\end{equation}
where $C_1$ and $C_2$ are constants. Naturally, these constants can be interpreted as a renormalization of the strength of the interaction between particles.

We formulate an improved mean-field model by analyzing more carefully the dependence on $S_m$ in Eq.~(\ref{langevinwb}). Following Refs.~\cite{pimpinelli,pimpinelli1}, we write

\begin{equation}
\left\langle \frac{S_m}{(x_{m+j+1}-x_{m+1})(x_{m+j+1}-x_m)}\right\rangle\rightarrow\varsigma_1(S),
\end{equation}
\begin{equation}
\left\langle \frac{S_m}{(x_{m}-x_{m-j})(x_{m+1}-x_{m-j})}\right\rangle\rightarrow\varsigma_1(S),
\end{equation}
where $\varsigma_1(S)=\tilde{C}_1\,S/(j(j+1)\left\langle S^2\right\rangle)$. Note that, in general, $\left\langle S^2\right\rangle$ is a function of $t$. However, in our mean-field approach the average of $S^2$ at time $t$ is replaced by the average in the stationary state $\left\langle\cdots \right\rangle_{\rm{st}}$. In order to simplify our notation, in the text we omit the subscript indicating stationary state. In the same way we write
\begin{equation}
\left\langle \frac{1}{x_{m+j+1}-x_{m+1}}\right\rangle=\left\langle \frac{1}{x_{m}-x_{m-j}}\right\rangle \rightarrow\varsigma_2(S),
\end{equation}
with $\varsigma_2(S)=\tilde{C}_1/(j\left\langle S\right\rangle)$. Finally, we use
\begin{equation}
\left\langle \frac{1}{S_m}\right\rangle\rightarrow\frac{\tilde{C}_2}{S}.
\end{equation}
Again, $\tilde{C}_1$ and $\tilde{C}_2$ are constants.

\section{Exact behavior of the Dyson's Brownian model for $s\ll1$}\label{app2}
From Eqs. (66) and (67) of Ref.~\cite{Bogomolny} one can find the behavior of $p^{(n;q)}(s)$ for small values of $s$. The case $n<q-1$ is given by Eq. (66) of Ref.~\cite{Bogomolny}. Since $\sum^{n+1}_i S_i\ll\left\langle S\right\rangle$, we find

\begin{equation}\label{app21}
P_{n+1}(S_1,\cdots,S_{n+1})\propto \int^{\infty}_0 dS_{n+2}\cdots \int^{\infty}_0 dS_{q-1}\zeta(\vec{S}) \vartheta(\vec{S}),
\end{equation}
where
\begin{equation}
\zeta(\vec{S})=\prod^{q-1}_{i=1}S^{\beta}_i \prod^{q-2}_{i=1}(S_i+S_{i+1})^{\beta} \cdots (S_i+\cdots+S_{q-1})^{\beta},
\end{equation}
and
\begin{equation}
\vartheta(\vec{S})=e^{-\frac{q\,\beta+1}{\left\langle S\right\rangle}\sum^{q-1}_{j=1}S_j}.
\end{equation}
After integration, Eq.~(\ref{app21}) reduces to
\begin{eqnarray}\label{eqap2a}
P_{n+1}(S_1,\cdots,S_{n+1})& \propto & \prod^{n+1}_{i=1}S^{\beta}_i \prod^{n}_{i=1}(S_i+S_{i+1})^{\beta}\times \cdots \nonumber \\
&\cdots & \times(S_1+\cdots+S_{n+1})^{\beta}\vartheta(\vec{S}).
\end{eqnarray}
Finally, using Eq.~(\ref{eqap2a}) with Eq.~(\ref{pnsdef}) we find
\begin{equation}
p^{(n;q)}(s)\propto s^{n+\sum^{n}_{i=0}(n+1-i)}\propto s^{\frac{\beta(n+1)(n+2)}{2}+n}.
\end{equation}
For $n\geq q-1$ we use Eq. (67) of Ref.~\cite{Bogomolny}. Then
\begin{equation}\label{eqapp6}
P_{n+1}(S_1,\cdots,S_{n+1})\propto \prod^{q-1}_{j=1} \prod^{n+1-j}_{i=1}(S_i+\cdots +S_{i+j})^{\beta}.
\end{equation}
Substituting Eq.~(\ref{eqapp6}) into Eq.~(\ref{pnsdef}) leads to
\begin{equation}
p^{(n;q)}(s)\propto s^{\beta \sum^{q-1}_{j=0}(n-j+1)+n}=s^{\frac{q}{2}(3+2n-q)\beta+n}.
\end{equation}


\begin{thebibliography}{0}
\bibitem{EWSGWS} In Ref.~\onlinecite{gonzalez2} this was called the generalized Wigner surmise, with the generalization being to $n>0$.  Here we call this feature ``extended", similar to Ref.~\onlinecite{abul}, to avoid confusion with the generalized Wigner surmise (GWS) discussed in Refs.~\onlinecite{pimpinelli,pimpinelli1,einstein,einstein1}, where the generalization is the value of $\beta$ to other than 1, 2, or 4.
\bibitem{por} Here, $p^{(n;\infty)}(s)$ is the probability that, given a particle at the center of a circle with radius $s$, there is another particle within an annulus centered at the first particle with internal radius $s$ and external radius $s+ds$. The condition that there are $n$ additional particles inside the disc of radius $s$ still applies.
\end{thebibliography}
\end{document}